\documentclass[a4paper]{article}

\usepackage{amsmath}
\usepackage[psamsfonts]{amssymb}
\usepackage{latexsym}
\usepackage{cmmib57}
\usepackage[dvips]{hyperref}

\usepackage{graphicx}
\usepackage{pstricks}
\usepackage{pst-coil}
\usepackage{pst-node}

\psset{unit=\unitlength}
\psset{arrowsize=2pt 5}

\newcommand{\goth}[1]{\mathfrak{#1}}
\newcommand{\scr}[1]{\mathcal{#1}}
\def\idty{{\leavevmode{\rm 1\ifmmode\mkern -4.8mu\else\kern -.3em\fi
      I}}}
\renewcommand{\Bbb}[1]{\if1#1\idty\else\mathbb{#1}\fi}

\newcommand{\kb}[1]{|#1\rangle\langle#1|}

\newcommand{\ket}[1]{|#1\rangle}

\newcommand{\tr}{\operatorname{tr}}

\newcommand{\Lz}{\operatorname{L}^2}

\newcommand{\Id}{\operatorname{Id}}
\newcommand{\Ad}{\operatorname{Ad}}

\newcommand{\restr}{\upharpoonright}
\newcommand{\HS}{{\rm HS}}

\newtheorem{thm}{Theorem}[section]
\newtheorem{defi}[thm]{Definition}
\newtheorem{prop}[thm]{Proposition}
\newtheorem{lem}[thm]{Lemma}
\newtheorem{kor}[thm]{Corollary}
\newenvironment{proof}{\par\noindent\textit{Proof.\ }}{\hfill $\Box$ \vspace{1em}}

\newtheorem{aX}{Axiom} 

\newcommand\calA{{\cal A}}

\newcommand\calK{{\cal K}}

\newcommand\CAR{{\cal A}^{\rm CAR}}

\newcommand\evl{\alpha_t}

\newcommand{\abs}[1]{\left|#1\right|}

\newcommand{\rbk}[1]{\left(#1\right)}

\newcommand{\cbk}[1]{\left\{#1\right\}}

\title{Entanglement, Haag-duality and type properties of infinite quantum spin chains}
 \author{ 
   M. Keyl\thanks{Electronic Mail: \tt{m.keyl@tu-bs.de}}\\[1ex]
   {\small Istituto Nazionale di Fisica della Materia, Unita' di Pavia,}\\
   {\small Dipartimento di Fisica ``A. Volta'', via Bassi 6, I-27100 Pavia, Italy}\\[1ex] 
   T. Matsui\thanks{Electronic Mail: \texttt{matsui@math.kyushu-u.ac.jp}}\\[1ex]
   {\small Graduate School of Mathematics, Kyushu Univ.,}\\
   {\small 1-10-6 Hakozaki, Fukuoka 812-8581, Japan}\\[1ex]
   D. Schlingemann\thanks{Electronic Mail: \texttt{d.schlingemann@tu-bs.de}} 
 {{}\ and\ } R.~F. Werner\thanks{Electronic Mail: \tt{r.werner@tu-bs.de}}
   \\[1ex]
  {\small Institut f{\"u}r Mathematische Physik, TU Braunschweig,}\\
  {\small Mendelssohnstr.3, 38106 Braunschweig, Germany.}}
\date{\today}

\begin{document}

\maketitle

\begin{abstract}
  We consider an infinite spin chain as a bipartite system consisting of the left and right half-chain and
  analyze entanglement properties of pure states with respect to this splitting. In this context we show that
  the amount of entanglement contained in a given state is deeply related to the von Neumann type of the
  observable algebras associated to the half-chains. Only the type I case belongs to the usual entanglement
  theory which deals with density operators on tensor product Hilbert spaces, and only in this situation
  separable normal states exist. In all other cases the corresponding state is infinitely entangled in the
  sense that one copy of the system in such a state is sufficient to distill an infinite amount of maximally
  entangled qubit pairs. We apply this results to the critical XY model and show that its unique ground state
  $\varphi_S$ provides a particular example for this type of entanglement.  
\end{abstract}

\section{Introduction}
\label{sec:introduction}

Entanglement theory is not only at the heart of quantum information theory, it has also produced a lot of very
deep (and in particular quantitative) insights into the structure of quantum correlations. The latter play
also a paramount role in condensed matter physics, in particular in the study of phase transitions and
critical phenomena. It is therefore an interesting and promising task to analyze how both fields can benefit
from each other, or in other words: to apply entanglement theory to models of quantum statistical mechanics.

A lot of research was recently done on this subject, concentrating in particular on one-dimensional systems 
(cf. \cite{AEPW02,MR2023564,MR2045771,BotRez04,MR2115123,MR2083138,MR2104889,Korepin04,MR2045771,%
  MR2115988,Eisert05a,MR2194026,MR2131351,MR2136600,OLEC05,WOVC05} and the references therein for a still
incomplete list). Many of these papers study a ground state of a spin chain model and calculate the von Neumann
entropy $S$ of its restriction to a finite, contiguous block. It turns out that the scaling behavior of $S$
with respect to the length $L$ of the block is intimately related to criticality: For critical models the entropy
$S(L)$ tends to diverge logarithmically (in the limit $L \to \infty$), while $\lim_{L \to \infty} S(L)$ remains finite in
the non-critical case. 

The relation of these results to entanglement theory is given by the fact that $S$ -- the \emph{entropy
  of entanglement} -- measures the rate of maximally entangled qubit pairs (``singlets''), which can be
distilled from \emph{an infinite supply of systems}, if only local operations and classical communication
(LOCC) are allowed. To be more precise consider a spin chain as a bipartite system consisting of a finite
block of length $L$ (given to Alice) and the rest (given to Bob), and assume that an infinite amount of 
chains is available. The entropy of entanglement $S(L)$ describes then the number of singlets Alice and Bob
can produce \emph{per chain},  if they are only allowed to communicate classically with each other and to
operate on their parts of the chains. While this is a natural concept for finite dimensional systems, it seems
to be odd for infinite degrees of freedoms, because we already have infinitely many systems. Hence it is more
natural to ask how many singlets Alice and Bob can produce (in terms of LOCC) \emph{if only one chain is
  available}. This question is discussed in \cite{Eisert05a,OLEC05}, and it turns out that in the critical
case this ``\emph{one-copy entanglement}'' diverges logarithmically as well (but with a smaller factor in
front of the logarithm). 

Let us change our point of view now slightly and consider a splitting of the chain into a left and right half,
rather than into a finite part and the rest. The results just discussed indicate that the one-copy
entanglement of a critical chain becomes infinite in this case. As shown in \cite{InfEnt} states of such a
type can not be described within the usual setup of entanglement theory (density operators on tensor product
Hilbert spaces) but require instead the application of operator algebraic methods. The purpose of the present
paper is to take this point of view seriously and to rediscuss entanglement properties of infinite quantum spin
chains in an appropriate (i.e. algebraic) mathematical context. The basic idea is to associate to each set $\Lambda$
of spins in the chain the C*-algebra $\scr{A}_\Lambda$ of observables localized in $\Lambda$, and to describe the systems
in term of this \emph{net of algebras} -- rather than in terms of a fixed Hilbert space. This is a well known
mathematical approach to quantum spin systems, and it has produced a lot of deep and powerful methods and
results (cf. the corresponding section of \cite{BraRob2} and the references therein). Of special importance
for us are the algebras $\scr{A}_L$ and $\scr{A}_R$ associated to the left ($L$) and right($R$)
half-chain. They represent the corresponding splitting of the spin chain into a bipartite system. In the
following we can think of $\scr{A}_L$ (respectively $\scr{A}_R$) as the algebra which is generated by the
observables available only to Alice (respectively Bob). The main message of this paper is now that the degree
of entanglement contained in a pure state of the chain is deeply related to properties (in particular the von
Neumann type) of the weak closure of $\scr{A}_L$ and $\scr{A}_R$ in the corresponding GNS representation. We
can show in particular that under mild technical assumption (most notably Haag-duality) two different cases
arise: 
\begin{itemize}
\item 
  \emph{The low entangled case,} where the half-chain algebras are of type I, separable normal states exist,
  but no normal state can have infinite one-copy entanglement. This covers the traditional setup of
  entanglement theory.
\item 
  \emph{The infinitely entangled case.} Here the half-chain algebras are not of type I, all normal states have
  infinite one-copy entanglement, and consequently no separable normal state exists. 
\end{itemize}
The previous results mentioned above indicate that critical models usually belong to the second case. Using
the method developed in \cite{MR743380,MR810491} we prove this conjecture explicitly for the critical XY
model. In this context we show in particular that the (unique) ground state of a critical XY chain satisfies
Haag duality.  

The outline of the paper is as follows: After presenting some notations and mathematical preliminaries in Section
\ref{sec:preliminaries} we will discuss (Section \ref{sec:entangl-c-algebr}) the generalizations of the usual
setup for entanglement theory which are necessary in a C*-algebraic context. This is mostly a review of
material presented elsewhere \cite{MR887998,InfEnt,MR2153773} adopted to the special needs of this paper. In
Section \ref{sec:entangl-von-neum} we analyze the relations between the von Neumann type of half-chain
algebras and the amount of entanglement in a given state (cf. the discussion in the last paragraph). These
results are then applied to spin chains. In Section \ref{sec:entangl-spin-chains} we treat kinematical
properties like translational invariance, localization of entanglement and cluster properties, while Section
\ref{sec:critical-xy-model} is devoted to a detailed study of the critical XY model. 

\section{Preliminaries}
\label{sec:preliminaries}

A quantum spin chain consists of infinitely many qubits (more generally $d$-level systems, but we are only 
interested in the spin 1/2 case) arranged on a one-dimensional regular lattice (i.e. $\Bbb{Z}$). We describe 
it in terms of the UHF $C^*-$algebra $2^{\infty}$ (the infinite tensor product of 2 by 2 matrix algebras ) : 
\begin{equation}
  {\scr{A}} = \overline{\bigotimes_{\Bbb{Z}} \: M_{2}(\Bbb{C})}^{C^*} . 
\end{equation}
Each component of the tensor product above is specified with a lattice site $\Bbb{Z}$. By $Q^{(j)}$ we denote
the element of ${\scr{A}}$ with $Q$ in the jth component of the tensor product and the identity in any other 
component. For a subset $\Lambda$ of $\Bbb{Z}$ , ${\scr{A}}_{\Lambda}$ is defined as the $C^*$-subalgebra of $\scr{A}$
generated by elements supported in  $\Lambda$. We set
\begin{equation} 
{\scr{A}}_{\rm loc} = \bigcup_{ \Lambda \subset {\Bbb{Z}} , \abs{\Lambda} < \infty}  \:\: \scr{A}_{\Lambda} 
\label{eqn:a1}
\end{equation}
where the cardinality of $\Lambda$ is denoted by $\abs{\Lambda}$. We call an element of ${\scr{A}}_{\Lambda}$ a local observable
or a strictly local observable. $\scr{A}_{\rm loc}$ is a dense subalgebra of $\scr{A}$.

In this paper we will look at spin chains as \emph{bipartite systems}. Hence we have to consider observables
and operations which are located on the right respectively left part of the chain.  They are described in
terms of the two half-chain algebras\footnote{We will use interval notations like $(a,b]$ frequently for
  subsets of $\Bbb{Z}$ rather than $\Bbb{R}$.}  
\begin{equation}
  \scr{A}_R = \scr{A}_{[0,\infty)},\quad \scr{A}_L = \scr{A}_{(-\infty,0)}.
\end{equation}
For each state $\omega$ of $\scr{A}$ we can introduce the restricted states 
\begin{equation}
  \omega_R = \omega \vert_{\scr{A}_R},\quad \omega_L = \omega \vert_{\scr{A}_L}
\end{equation}
and the von Neumann algebras
\begin{equation}
  \scr{R}_{R,\omega} = \pi_\omega(\scr{A}_R)'',\quad \scr{R}_{L,\omega} = \pi_\omega(\scr{A}_L)'',
\end{equation}
where $(\scr{H}_\omega,\pi_\omega,\Omega_\omega)$ denotes the GNS representation associated with $\omega$. 
For arbitrary $\omega$ the two von Neumann algebras $\scr{R}_{L/R,\omega}$ have the following properties: 
\begin{itemize}
\item 
  Since $\scr{A}_{L/R}$ are generated (as C*-algebras) by an increasing sequence of finite dimensional matrix
  algebras, the same holds for $\scr{R}_{R/L,\omega}$ in the weak topology. Hence the $\scr{R}_{L/R,\omega}$ are
  \emph{hyperfinite}.
\item 
  For the same reason the GNS Hilbert space $\scr{H}_\omega$ is separable, hence $\scr{R}_{L,\omega}$ and
  $\scr{R}_{R,\omega}$ are \emph{$\sigma$-finite}.
\item 
  $\scr{R}_{L,\omega}$ and $\scr{R}_{R,\omega}$ are \emph{mutually commuting}, i.e. $[A,B]=0$ for all $A \in
  \scr{R}_{L,\omega}$ and all $B \in \scr{R}_{R,\omega}$. 
\end{itemize}
If $\omega$ is \emph{pure}, the GNS representation $\pi_\omega$ is irreducible and we get in addition:
\begin{itemize}
\item 
  $\scr{R}_{L,\omega}$ and $\scr{R}_{R,\omega}$ together \emph{generate} $\pi_\omega(\scr{A})'' = \scr{B}(\scr{H}_\omega)$, i.e. 
    \begin{equation} \label{eq:7}
    \scr{R}_{L,\omega} \lor \scr{R}_{R,\omega} = (\scr{R}_{L,\omega} \cup \scr{R}_{R,\omega})'' = \scr{B}(\scr{H}_\omega).
  \end{equation}
\item 
  The $\scr{R}_{L/R,\omega}$ are \emph{factors}. This can  be seen as follows: The center $\scr{Z}_\omega$ of
  $\scr{R}_{L,\omega}$ satisfies 
  \begin{equation} 
    \scr{Z}_\omega' = (\scr{R}_{L,\omega} \cap \scr{R}_{L,\omega}')' = \scr{R}_{L,\omega} \lor \scr{R}_{L,\omega}'.
  \end{equation}
  But $\scr{R}_{R,\omega} \subset \scr{R}_{L,\omega}'$ and from Equation (\ref{eq:7}) we therefore get $\scr{Z}_\omega' =
  \scr{B}(\scr{H}_\omega)$. Hence $\scr{R}_{L,\omega}$ is a factor, and $\scr{R}_{R,\omega}$ can be treated similarly.
\end{itemize}

A special class of states we will consider frequently are \emph{translationally invariant states}. A state $\omega$
is translationally invariant if $\omega \circ \tau_1 = \omega$ holds, where $\tau_1$ denotes the automorphism which shifts the
whole chain one step to the right. More precisely, we define for each $k \in \Bbb{Z}$ an automorphism $\tau_k$ of
$\scr{A}$ by $ \tau_j(Q^{(k)})=Q^{(j+k)}$ for any $j \in \Bbb{Z}$ and any $2 \times 2$ matrix $Q$. 

A particular example of a translationally invariant state is the \emph{ground state of the
  critical XY-model}. To give its definition, note first that a state $\varphi$ is a \emph{ground state} with
respect to a one parameter group $\evl$ of automorphisms of $\scr{A}$, if   
\begin{equation}
\varphi (Q^* \delta (Q))  \geq 0 
\label{eqn:a2}
\end{equation}
holds for any $Q$ in the domain of the generator $\delta$ of $\evl$, where
\begin{equation}
  \delta (Q) = -i \frac{d}{dt}\evl (Q) \vert_{t=0} . 
\end{equation}
The dynamics of the XY-model is given formally by 
\begin{equation}
  \evl (Q) =  \exp^{it H_{XY}} Q \exp^{-it H_{XY}}, \quad Q \in \scr{A} 
\end{equation}
with the Hamiltonian
\begin{equation}
  H_{XY} = -\sum_{j \in \Bbb{Z}} \{ (1+\gamma)\sigma_x^{(j)}\sigma_x^{(j+1)} +(1-\gamma)\sigma_y^{(j)}\sigma_y^{(j+1)}  +2\lambda \sigma_z^{(j)}\},
\end{equation}
where $\sigma_x^{(j)}$ , $\sigma_y^{(j)}$, and $\sigma_z^{(j)}$ are Pauli spin matrices at the site $j$ and $\gamma$ and $\lambda$ are
real parameters (anisotropy and magnetic field). The precise mathematical definition of $\evl$ is obtained via
thermodynamic limit: If we set 
\begin{equation}
H_{XY} ([a,b])= -\sum_{j=a}^{b-1 } \{ (1+\gamma)\sigma_x^{(j)}\sigma_x^{(j+1)} +(1-\gamma)\sigma_y^{(j)}\sigma_y^{(j+1)}  +2\lambda \sigma_z^{(j)}\},
\label{eqn:a3}
\end{equation}
the limit
\begin{equation} \label{eq:5}
  \evl (Q) = \lim_{N \to \infty} e^{ it H_{XY}([-N,N])} Q e^{-it H_{XY}([-N,N])},\quad Q \in \scr{A}
\end{equation}
 exists in norm topology of $\scr{A}$ and defines the time evolution $\evl$. The local algebra $\scr{A}_{\rm
   loc}$ is a core for the generator $\delta(Q) = [H_{XY}, Q]$.  

The \emph{critical} XY model arises if the parameter $\lambda,\gamma$ satisfy $\abs{\lambda} =1 , \gamma \neq 0$ or $\abs{\lambda} <1 , \gamma =
0$. In this case it is known (cf. Theorem 1 of \cite{MR810491}) that there is a unique ground state $\varphi_S$. We will
refer to it throughout this paper as ``the (unique) ground state of the critical XY model''. 

\section{Entanglement and C*-algebras}
\label{sec:entangl-c-algebr}

Our aim is to look at an infinite quantum spin chain as a bipartite system which consists of the left and
right half-chain and to analyze entanglement properties which are related to this splitting. However, in our
model the two halfs of the chain are not described by different tensor factors of a tensor product Hilbert
space, but by different subalgebras of the quasi-local algebra $\scr{A}$. Therefore we have to generalize some
concepts of entanglement theory accordingly (cf. also \cite{MR887998,InfEnt,MR2153773}). 

\begin{defi}
  A \emph{bipartite system} is a pair of unital C*-algebras $\goth{A}$, $\goth{B}$ which are both subalgebras
  of the same ``ambient algebra'' $\goth{M}$, commute elementwise ($[A,B] = 0$ for all $A \in \goth{A}$, $B \in
  \goth{B}$) and satisfy $\goth{A} \cap \goth{B} = \Bbb{C}\Bbb{1}$. 
\end{defi}

For the spin chain we have $\goth{A} = \scr{A}_L$, $\goth{B} = \scr{A}_R$ and $\goth{M} = \scr{A}$. The usual
setup in terms of a tensor product Hilbert space $\scr{H}_1 \otimes \scr{H}_2$ arises with $\scr{A} =
\scr{B}(\scr{H}_1) \otimes \Bbb{1}$, $\goth{B} = \Bbb{1} \otimes \scr{B}(\scr{H}_2)$ and $\goth{M} = \scr{B}(\scr{H}_1 \otimes
\scr{H}_2)$; we will refer to this situation as the ``type I case'' (since $\goth{A}$ and $\goth{B}$ are type
I von Neumann algebras in this case). If $\goth{M}$ is finite dimensional, the latter is the only possible
realization of bipartite systems -- in full compliance with ordinary (i.e. finite dimensional) entanglement
theory.   

\begin{defi}
  A state $\omega$ on the ambient algebra is called a \emph{product state} if $\omega(AB) = \omega(A)\omega(B)$ for all $A \in
  \goth{A}$, $B \in \goth{B}$; i.e. if $\omega$ does not contain any correlations. $\omega$ is \emph{separable} if it is
  an element of the weakly closed convex hull of the set of product states. If $\omega$ is not separable, it is
  called \emph{entangled}.    
\end{defi}

If $\omega$ is a normal state of a type I system, i.e. $\omega(A) = \tr(\rho A)$ with a density operator $\rho$ on $\scr{H}_1 \otimes
\scr{H}_2$ we see immediately that $\omega$ is a product state iff $\rho=\rho_1 \otimes \rho_2$ holds. Hence we recover the usual
definitions. 

Given a bipartite system in an entangled state our aim is to extract maximally entangled qubit pairs using
an operation $T$  which does not generate entanglement itself (i.e. $T$ should map separable states to separable
states -- such a map is called \emph{separable} itself). For the purpose of this paper it is sufficient, to
look only at the most simple class of such maps: local operations (for LOCC maps cf. \cite{MR2153773}).

\begin{defi} \label{def:2}
  A \emph{local operation} between two bipartite systems $\goth{A}_1, \goth{B}_1 \subset \goth{M}_1$ and $\goth{A}_2,
  \goth{B}_2 \subset \goth{M}_2$ is a unital completely positive (cp) map $T: \goth{M}_1 \to \goth{M}_2$ such that
  \begin{enumerate}
  \item \label{item:1}
    $T(\goth{A}_1) \subset \goth{A}_2$ and $T(\goth{B}_1) \subset \goth{B}_2$,
  \item \label{item:2}
    and $T(AB) = T(A)T(B)$ holds for all $A \in \goth{A}_1$, $B \in \goth{B}_1$.  
  \end{enumerate}
\end{defi}

If we consider in the type I case an operation $T: \goth{A}_1 \otimes \goth{B}_1 \to \goth{A}_2 \otimes \goth{B}_2$ which is
local and \emph{normal}, it must have the form $T=T_1\otimes T_2$ with two unital (and normal) cp maps $T_1$,
$T_2$. To see this, expand an element $Q \in \goth{M}_1 = \goth{A}_1 \otimes \goth{B}_1 = \scr{B}(\scr{H}_1 \otimes
\scr{H}_2)$ in terms of matrix units $e_{i,j}$. By normality we get
\begin{equation}
   Q = \sum_{ijkl}  c_{ijkl} e_{ij}\otimes e_{kl} \: , \: T(Q)=\sum_{ijkl}  c_{ijkl} T(e_{ij}\otimes 1) T(1 \otimes e_{kl} ) = T_1 \otimes T_2 (Q), 
\end{equation}
hence $T = T_1 \otimes T_2$ as stated. Note that $T$ would not factorize if we consider only item \ref{item:1} of
this  definition: If $\omega$ is a state on $\goth{M}_1$ the map $T(A) = \Bbb{1} \omega(A)$ satisfies condition
\ref{item:1} even if $\omega$ is entangled. To fulfill condition \ref{item:2} as well, however, $\omega$ has to be a
product state.  

Usual distillation protocols describe procedures, to extract a certain amount of entanglement \emph{per
  system}, if a large (possibly infinite) number of equally prepared systems is available. However, if we
study an infinite quantum spin chain, we have already a system consisting of infinitely many particles. Hence
one copy of the chain could be sufficient for distillation purposes, and if the total amount of entanglement
contained in the system is infinite, it might be even possible to extract infinitely many singlets from
it. This idea is the motivation for the following definition\footnote{Note that the definition given in
  \cite{Eisert05a} is slightly different from ours, because the condition $T^*(\omega) = \kb{\chi_d}$ is used instead
  of Equation (\ref{eq:1}). The advantage of our approach (following \cite{InfEnt}) lies in the fact that
  topological questions concerning the limit $\epsilon\to0$ can be avoided.} \cite{InfEnt,Eisert05a}. 

\begin{defi} \label{def:1}
  Consider a state $\omega$ of a bipartite system $\goth{A}, \goth{B} \subset \goth{M}$. The quantity $E_1(\omega) =
  \log_2(d)$ is called the \emph{one copy entanglement} of $\omega$ (with respect to $\goth{A}$, $\goth{B}$), if
  $d$ is the biggest integer $d\geq2$ which admit for each $\epsilon>0$ a local operation  $T_\epsilon:
  \scr{B}(\Bbb{C}^d) \otimes \scr{B}(\Bbb{C}^d) \to \goth{M}$ such that 
  \begin{equation} \label{eq:1}
    \omega\bigl(T_\epsilon(\kb{\chi_d})\bigr) > 1 - \epsilon,\quad \chi_d = \frac{1}{\sqrt{d}} \sum_{j=1}^d \ket{jj}
  \end{equation}
  holds. If no such $d$ exists we set $E_1(\omega) = 0$ and if (\ref{eq:1}) holds for all $d\geq2$ we say that $\omega$
  contains \emph{infinite one copy entanglement} (i.e. $E_1(\omega) = \infty$).
\end{defi}

The next result is a technical lemma which we will need later on (cf. \cite{MR2153773} for a proof). It allows
us to transfer results we have got for C*-algebras $\goth{A}, \goth{B}$ to the enveloping von Neumann
algebras $\goth{A}'', \goth{B}''$ and vice versa. 

\begin{lem} \label{lem:1}
  Consider a bipartite system $\goth{A}, \goth{B} \subset \goth{M} \subset \scr{B}(\scr{H})$ with irreducible $\goth{M}$
  and a density operator $\rho$ on $\scr{H}$. The state $\tr(\rho \,\cdot\,)$ has infinite one copy entanglement with
  respect to $\goth{A}$, $\goth{B}$ iff the same is true with respect to $\goth{A}''$, $\goth{B}''$.
\end{lem}

Finally, we will consider violations of Bell inequalities. This subject is studied within an algebraic context
in \cite{MR887998}. Following these papers let us define:

\begin{defi}
  Consider a bipartite system $\goth{A}, \goth{B} \subset \goth{M}$. The \emph{Bell correlations} in a state $\omega:
  \goth{M} \to \Bbb{C}$ are defined by
  \begin{equation}
    \beta(\omega) = \frac{1}{2} \sup \omega(A_1(B_1+B_2)+A_2(B_1-B_2)),
  \end{equation}
  where the supremum is taken over all selfadjoint $A_i \in \goth{A}$, $B_j \in \goth{B}$ satisfying
  $-\idty \leq A_i \leq \idty$, $-\idty \leq B_j \leq \idty$, for $i,j=1,2$. In other words $A_1,A_2$ and
  $B_1,B_2$ are (appropriately bounded) observables measurable by Alice respectively Bob. 
\end{defi}

Of course, a classically correlated (separable) state, or any other state consistent with a local hidden
variable model \cite{Werner89} satisfies the Bell-CHSH-inequality $\beta(\omega)\leq1$, while any $\omega$ has to satisfy Cirelson's
inequality \cite{Cirelson,SumWer95,BEG} 
\begin{equation}
  \beta(\omega) \leq \sqrt{2}.
\end{equation}
If the upper bound $\sqrt{2}$ is attained we speak of a \emph{maximal violation} of Bell's
inequality.

\section{Entanglement and von Neumann type}
\label{sec:entangl-von-neum}

In this section we want to consider the special case that $\goth{A}$ and $\goth{B}$ are von Neumann algebras
acting on a Hilbert space $\scr{H}$ and having all the properties mentioned in Section
\ref{sec:preliminaries}. In other words: $\goth{A}$ and $\goth{B}$ are hyperfinite and $\sigma$-finite factors, and
they generate together $\scr{B}(\scr{H})$, i.e.
\begin{equation} \label{eq:8}
  \goth{A} \lor \goth{B} = \scr{B}(\scr{H}).
\end{equation}
As the ambient algebra we choose $\goth{M} = \scr{B}(\scr{H})$ and we will call a bipartite system with these
properties in the following \emph{simple}. If in addition $\goth{A}' = \goth{B}$ holds we say that \emph{Haag
  duality} holds. We will see that these conditions are already quite  restrictive (in particular Equation
(\ref{eq:8})) and lead to a close relation between entanglement and the type of the factors $\goth{A}$ and
$\goth{B}$.  

\subsection{Split property}
\label{sec:split-properties}

Let us consider first the \emph{low entangled case}. It is best characterized by the \emph{split property}, 
i.e. there is a type I factor $\scr{N}$ such that
\begin{equation} \label{eq:6}
  \goth{A} \subset \scr{N} \subset \goth{B}'
\end{equation}
holds. In this case normal states with infinite one copy entanglement does not exist. More precisely we have
the following theorem.

\begin{thm} \label{thm:4}
  Consider a simple bipartite system $\goth{A}, \goth{B} \subset \scr{B}(\scr{H})$ satisfying the split property
  (\ref{eq:6}). Then there is no normal state on $\scr{B}(\scr{H})$ with infinite one copy entanglement.
\end{thm}

The proof of this theorem can be divided into two steps. The first one shows that the split property forces
the algebras $\goth{A}, \goth{B}$ to be of type I.

\begin{prop} \label{prop:1}
  A simple bipartite system $\goth{A}, \goth{B} \subset \scr{B}(\scr{H})$ satisfies the split property iff it is (up
  to unitary equivalence) of the form $\scr{H} = \scr{H}_1 \otimes \scr{H}_2$, $\goth{A} = \scr{B}(\scr{H}_1) \otimes
  \Bbb{1}$ and $\goth{B} = \Bbb{1} \otimes \scr{B}(\scr{H}_2)$. This shows in particular that the split property
  implies Haag duality.  
\end{prop}

\begin{proof}
  If $\goth{A}, \goth{B}$ are of the given form, the split property holds trivially with $\scr{N} =
  \goth{A}$. Hence only the other implications has to be proved. To this end consider the relative commutant
  $\scr{M} = \goth{A}' \cap \scr{N}$ of $\goth{A}$ in $\scr{N}$. Since $\scr{N} \subset \goth{B}'$ we have $\scr{M} \subset
  \goth{A}'$ and $\scr{M} \subset \goth{B}'$. Hence with Equation (\ref{eq:8}) 
  \begin{equation} \label{eq:24}
    \scr{M} \subset (\goth{A} \lor \goth{B})' = \Bbb{C} \Bbb{1}.
  \end{equation}
  Since $\scr{N}$ is of type I, there are Hilbert spaces $\scr{H}_1, \scr{H}_2$ and a unitary $U: \scr{H} \to
  \scr{H}_1 \otimes \scr{H}_2$ such that $U \scr{N} U^* = \scr{B}(\scr{H}_1) \otimes \Bbb{1}$ holds
  \cite[Thm. V.1.31]{MR548728}. Hence $\goth{A} \subset \scr{N}$ implies $U\goth{A}U^* = \tilde{\goth{A}} \otimes
  \Bbb{1}$, with a subalgebra $\tilde{\goth{A}}$ of $\scr{B}(\scr{H}_1)$. Equation (\ref{eq:24}) therefore
  leads to $\tilde{\goth{A}}' = \Bbb{C} \Bbb{1}$; hence $\tilde{\goth{A}} = \scr{B}(\scr{H}_1)$ and $U
  \goth{A} U^* = \scr{B}(\scr{H}_1) \otimes \Bbb{1}$ as stated. In a similar way we can show that $U\goth{B}U^* =
  \Bbb{1} \otimes \scr{B}(\scr{H}_2)$, which concludes the proof.
\end{proof}

Roughly speaking we can say that there is not enough room between $\goth{A}$ and $\goth{B}'$ to allow
non-trivial splits with $\goth{A} \neq \scr{N}$. This is exactly the converse of a \emph{standard} split
inclusion, where $\goth{A}' \cap \goth{B}'$ is big enough to admit a cyclic vector \cite{MR735338,MR739630}. 

With this proposition Theorem \ref{thm:4} follows immediately from a recent result about the type I case
\cite{InfEnt}:   

\begin{prop} \label{prop:5}
  Consider a normal state $\omega$ of a type I bipartite system ($\goth{A} = \scr{B}(\scr{H}_A) \otimes \Bbb{1}$,
  $\goth{B} = \Bbb{1} \otimes \goth{B}(\scr{H}_B) \subset \goth{M} = \scr{B}(\scr{H}_A \otimes \scr{H}_B)$). For each sequence
  of unital cp-maps 
  \[T_d : \scr{B}(\Bbb{C}^d \otimes \Bbb{C}^d) \to \goth{M} \]
  such that $T_d^*\phi$ \emph{is ppt}\footnote{I.e. the density operator associated to $T^*\phi$ has positive
    partial transpose.} \emph{for each pure product state} $\phi$, we have  
  \[ \lim_{d \to \infty} \omega\bigl(T_d(\kb{\chi_d})\bigr) = 0,\quad \chi_d = \frac{1}{\sqrt{d}} \sum_{j=1}^d \ket{jj}. \]  
\end{prop}

The operations $T_d$ considered here map pure product states to ppt-states. This is a much weaker condition
than separability (and therefore much weaker than LOCC). Hence this theorem covers all physically relevant
variations of Definition \ref{def:1}. Note in addition that the possibility of normal states with
\emph{infinite distillable entanglement} is not excluded, because the usual entanglement distillation allows
the usage of an infinite supply of systems not just one copy. It is in fact easy to see that in type-I systems
with $\dim \scr{H}_A =\dim \scr{H}_B= \infty$ normal states with infinite distillable entanglement are in a certain
sense generic (cf. \cite{CliftHalv99,HorCirLew} for details).   

The result of this subsection shows that the split property (\ref{eq:6}) characterizes exactly the traditional
setup of entanglement theory. Hence there are normal states which are separable but no normal state has
infinite one copy entanglement. This is the reason why we have called this case the ``low entangled'' one. 

\subsection{The maximally entangled case}
\label{sec:maxim-entangl-case}

The prototype of a state with infinite one-copy entanglement is a system consisting of infinitely many qubit
pairs, each in a maximally entangled state. It can be realized on a  spin chain as follows: Consider the
algebra $\scr{A}_{\{-j,j-1\}}$ containing all observables localized at lattice sites $-j$ and $j-1$. It is
naturally isomorphic to $\scr{B}(\Bbb{C}^2) \otimes \scr{B}(\Bbb{C}^2)$. Therefore we can define the state 
\begin{equation} \label{eq:20}
  \omega^{\{-j,j-1\}}_1(A) = \tr\bigl(\kb{\chi_2} A)
\end{equation}
with $\chi_2$ from Equation (\ref{eq:1}). It represents a maximally entangled state between the qubits at site
$-j$ and $j-1$. Now we can consider the infinite tensor product
\begin{equation} \label{eq:4a}
  \omega_1 = \bigotimes_{j \in \Bbb{N}} \omega^{\{-j,j+1\}}_1,
\end{equation}
which has obviously infinite one-copy entanglement. In \cite{InfEnt} it is argued that this state is the natural
analog of a maximally entangled state in infinite dimensions.

The left and right half-chain von Neumann algebras\footnote{To avoid clumsy notations we will write
  occasionally $\scr{H}_1$ etc. instead of $\scr{H}_{\omega_1}$, i.e. we will replace double indices $\omega_j$ by an
  index $j$.} $\scr{R}_{L,1}$ and $\scr{R}_{R,1}$ have the following properties \cite{InfEnt}
\begin{itemize}
\item 
  $\scr{R}_{L,1}, \scr{R}_{R,1} \subset \scr{B}(\scr{H}_1)$ form a simple bipartite system.
\item 
  Haag duality  holds: $\scr{R}_{R,1} = \scr{R}_{L,1}'$.
\item 
  $\scr{R}_{L,1}$ and $\scr{R}_{R,1}$ are hyperfinite type II$_1$ factors.
\end{itemize}
Note that the last property can be seen very easily, because the construction shown in the last paragraph is
exactly the Araki-Woods construction of the hyperfinite type II$_1$ factor (\cite{AraWoo}, cf. also
\cite[Thm. 2]{InfEnt} for a direct proof of the type II$_1$ property). Since all hyperfinite type II$_1$
factors are mutually isomorphic the maximally entangled case can be characterized as follows:

\begin{prop} \label{prop:6}
  Consider a hyperfinite type II$_1$ factor $\scr{M} \subset \scr{B}(\scr{H})$ admitting a cyclic and separating
  vector. Then the following statements hold:
  \begin{enumerate}
  \item \label{item:6}
    The pair $\scr{M}, \scr{M}' \subset \scr{B}(\scr{H})$ defines a simple bipartite system which is unitarily
    equivalent to $\scr{R}_{L,1}, \scr{R}_{R,1} \subset \scr{B}(\scr{H}_1)$.
  \item \label{item:10}
    Each normal state on $\scr{B}(\scr{H})$ has infinite one-copy entanglement (with respect to $\scr{M},
    \scr{M}'$). 
  \end{enumerate}
\end{prop}

\begin{proof}
  Since $\scr{M}$ and $\scr{R}_{L,1}$ are hyperfinite type II$_1$ factors, they are
  isomorphic \cite[Thm XIV.2.4]{MR1943007} and since both have a cyclic and separating vector this isomorphism
  is implemented by a unitary $U$. Hence $U^* \scr{M} U = \scr{R}_{L,1}$ and due to $\scr{R}_{R,1} =
  \scr{R}_{L,1}'$ \cite{InfEnt} we also have $U \scr{M}' U^* = \scr{R}_{R,1}$. This already proves item \ref{item:6}.

  To prove item \ref{item:10} it is sufficient to show the statement for $\scr{R}_{L,1}, \scr{R}_{R,1}$ rather
  than a general pair $\scr{M}, \scr{M}'$. Hence consider a density matrix $\rho$ on $\scr{H}_1$ and the corresponding
  state $\omega(A) = \tr\bigl(\rho \pi_1(A)\bigr)$ on the quasi-local algebra $\scr{A}$. According to Lemma \ref{lem:1},
  $\rho$ has infinite one copy  entanglement with respect to $\scr{R}_{L,1}, \scr{R}_{R,1}$ iff $\omega$ has infinite
  one copy entanglement with respect to $\scr{A}_L$, $\scr{A}_R$. Therefore, it is sufficient to prove the latter.

  To this end, note first that $\omega_1$ is pure and $\pi_1$ therefore irreducible. If $\rho = \kb{\psi}$ with a
  normalized $\psi \in \scr{H}_1$ this implies that $\omega(A) = \langle\psi,\pi_1(A)\psi\rangle$ is pure (in particular factorial) and
  unitarily equivalent to $\omega_1$. Hence we can apply Corollary 2.6.11 of \cite{BraRob1} which shows that
  quasi-equivalence of $\omega$ and $\omega_1$ implies that for each $\epsilon>0$ there is an $N \in \Bbb{N}$ 
  with   
  \begin{equation} \label{eq:2}
    |\omega(A) - \omega_1(A)| < \epsilon \|A\| \quad \forall A \in \scr{A}_{\{|n| > N\}}. 
  \end{equation}

  Now assume that $\rho$ is a general density matrix and $\omega$ therefore a mixed normal state on $\scr{A}$. If the
  spectral decomposition of $\rho$ is $\rho = \sum_j \lambda_j \kb{\psi_j}$ we have for each $\epsilon>0$ a $J \in \Bbb{N}$ with 
  \begin{equation}
    \|\omega - \omega_J\| < \frac{\epsilon}{3}\ \text{and}\ \omega_J(A) = \sum_{j=1}^J \lambda_j \omega_j(A) = \sum_{j=1}^J \lambda_j \langle\psi_j, \pi_1(A) \psi_j\rangle.
  \end{equation}
  The $\omega_j$ are pure states. Hence we find as in Equation (\ref{eq:2}) an $N \in \Bbb{N}$ such that
  \begin{equation}
    |\omega_j(A) - \omega_1(A)| <\frac{\epsilon}{3J} \|A\| \quad \forall A \in \scr{A}_{\{|n|>N\}}\ \forall j=1, \ldots, J 
  \end{equation}
  holds. By construction we have in addition $\left|1 - \Sigma_{j=1}^J \lambda_j\right| < \epsilon/3$. Therefore we get for all
  $A \in \scr{A}_{\{|n|  > N\}}$ with $\|A\|=1$:
  \begin{align}
    |\omega(A) - \omega_1(A)| &\leq |\omega(A) - \omega_J(A)| + |\omega_J(A) - \omega_1(A)| \\
    &\leq \frac{\epsilon}{3} + \sum_{j=1}^J \lambda_j | \omega_j(A) - \omega_1(A)| + \left|1 - \sum_{j=1}^J \lambda_j\right| |\omega_1(A)| \leq \epsilon. \label{eq:3}
  \end{align}

  Now consider the natural isomorphism 
  \begin{equation}
    T_{NM} : \scr{B}(\Bbb{C}^{2M} \otimes \Bbb{C}^{2M}) \to \scr{A}_{[-N-M,-N] \cup [N-1,N+M-1]} \subset \scr{A}.
  \end{equation}
  It satisfies by construction $\omega_1(T_{NM} \chi_2^{\otimes M}) = 1$. Together with Equation (\ref{eq:3}) this implies 
  (with $\| T_{NM} \chi_2^{\otimes M}\| =1$ since $\chi_2^{\otimes M}$ is a projector)
  \begin{align}
    |\omega(T_{NM} \chi_2^{\otimes M})| &\geq |\omega_1(T_{NM} \chi_2^{\otimes M})| - |\omega_1(T_{NM} \chi_2^{\otimes M}) - \omega(T_{NM} \chi_2^{\otimes M})| \\
    &\geq 1 -\epsilon \| T_{NM} \chi_2^{\otimes M}\| = 1 - \epsilon,
  \end{align}
  which shows that $\omega$ has infinite one copy entanglement.
\end{proof}

The bipartite systems described in this proposition admit only normal states which have infinite one-copy
entanglement. Hence there are in particular no normal, separable states. This is exactly the converse of the
split situation described in the last subsection, and we can call it: ``the maximally entangled case''. 

\subsection{Haag duality}
\label{sec:haag-duality}

Let us consider now simple bipartite systems which are not split but satisfy Haag duality. Then we always can
extract a maximally entangled system (as described in the last subsection) in terms of a local operation. 

\begin{prop} \label{prop:9}
  Consider a simple bipartite system $\goth{A}, \goth{B} = \goth{A}' \subset \scr{B}(\scr{H})$ such that $\goth{A}$
  is not of type I. Then there is an operation $\gamma: \scr{B}(\scr{H}_1) \to \scr{B}(\scr{H})$ which is local with
  respect to $\scr{R}_{L/R,1}$ and $\goth{A}, \goth{B}$.  
\end{prop}

\begin{proof}
  By assumption $\goth{A}$ is a factor, not of type I and $\goth{B} = \goth{A}'$. Hence $\goth{A}, \goth{B}$
  are either both of type II or both of type III. 

  If $\goth{A}$ and $\goth{B}$ are \emph{type II$_\infty$}, let us define the additional von Neumann algebras
  \begin{equation} \label{eq:17}
    \scr{M}_L = \scr{B}(\scr{H}_L) \otimes \scr{R}_{L,1} \otimes \Bbb{1}_R,\quad \scr{M}_L' = \scr{M}_R = \Bbb{1}_R \otimes
    \scr{R}_{R,1} \otimes \scr{B}(\scr{H}_R),
  \end{equation}
  where $\scr{H}_{L/R}$ are two infinite dimensional, separable Hilbert spaces and $\Bbb{1}_{L/R}$ are the unit
  operators on them. Since $\scr{R}_{L/R,1}$ are hyperfinite type II$_1$ factors, the $\scr{M}_{L/R}$ are
  hyperfinite type II$_\infty$ factors satisfying $\scr{M}_L' = \scr{M}_R$. By assumption the same is true for
  $\goth{A}$, $\goth{B}$. Hence there is a *-isomorphism $\gamma: \scr{M}_L \to \goth{A}$ (since the hyperfinite type
  II$_\infty$ factor is unique up to isomorphism \cite{MR1943007}).      

  Since $\goth{A}$, $\scr{M}_L$ and their commutants are $\sigma$-finite, purely infinite factors both admit a
  cyclic and separating vector \cite[Prop. 9.1.6]{MR1468230}. Hence the isomorphism $\gamma$ is unitarily
  implemented \cite[Thm 7.2.9]{MR1468230}, i.e. $\gamma(A) = UAU^*$ with a unitary $U : \scr{H}_L \otimes \scr{H}_1 \otimes
  \scr{H}_R \to \scr{H}$. Since   
  \begin{equation} \label{eq:4}
      U\scr{M}_LU^* = \goth{A}\ \text{and}\ U\scr{M}_RU^* = U\scr{M}_L'U^* = \goth{A}' = \goth{B}
  \end{equation}
  we get a local operation (even a local *-homomorphism) by
  \begin{equation}
    \scr{B}(\scr{H}_1) \ni A \mapsto U(\Bbb{1}_L \otimes A \otimes \Bbb{1}_R)U^* \in \scr{B}(\scr{H}),
  \end{equation}
  which proves the statement in the type II$_\infty$ case (note that Haag duality entered in Equation (\ref{eq:4})).

  If $\goth{A}$ and $\goth{B}$ are \emph{both of type II$_1$} we can define in analogy to Equation
  (\ref{eq:17}) the hyperfinite II$_\infty$   factors
  \begin{equation}
    \goth{A}_1 =  \scr{B}(\scr{H}_L) \otimes \goth{A} \otimes \Bbb{1}_R,\quad \goth{B}_1 = \Bbb{1}_L \otimes \goth{B} \otimes
    \scr{B}(\scr{H}_R) 
  \end{equation}
  As in the previous paragraph there exists a unitary $U: \scr{H}_L \otimes \scr{H}_1 \otimes \scr{H}_R \to \scr{H}_L \otimes
  \scr{H} \otimes \scr{H}_R$ such that Equation (\ref{eq:4}) holds with $\goth{A}, \goth{B}$ replaced by
  $\goth{A}_1, \goth{B}_1$. Hence with the density matrices $\rho_L$ on $\scr{H}_L$ and $\rho_R$ on $\scr{H}_R$ we
  can define a local operation $\scr{B}(\scr{H}_1) \to \scr{B}(\scr{H})$ by
  \begin{equation}
    \scr{B}(\scr{H}_1) \ni A \mapsto \tr_{LR}\bigl( \rho_L \otimes \Bbb{1}  \otimes \rho_R U(\Bbb{1}_L \otimes A  \otimes \Bbb{1}_R)U^* \bigr) \in
    \scr{B}(\scr{H}), 
  \end{equation}
  where $\tr_{LR}$ denotes the partial trace over $\scr{H}_L \otimes \scr{H}_R$. 

  If \emph{one algebra is type II$_\infty$} and \emph{the other type II$_1$} we can proceed in the same way, if we 
  adjoin only one type I factor to $\scr{B}(\scr{H})$, i.e. either $\scr{B}(\scr{H}_L)$ or
  $\scr{B}(\scr{H}_R)$.  

  Hence only the \emph{type III case} remains. If $\goth{A}$ is a hyperfinite type III factor it is
  \emph{strongly stable} (cf. Appendix \ref{sec:strong-stab-hyperf}), i.e.
  \begin{equation}
    \goth{A} \cong \goth{A} \otimes \scr{R}_{L,1}
  \end{equation}
  holds. By the same argument which let to Equation (\ref{eq:4}) this implies the existence of a unitary $U:
   \scr{H} \otimes \scr{H}_1 \to \scr{H}$ such that 
   \begin{equation} \label{eq:10}
     U \goth{A} \otimes \scr{R}_{L,1} U^* = \goth{A},\ \text{and}\ U \goth{B} \otimes \scr{R}_{R,1} U^* = \goth{B}.
   \end{equation}
   Therefore the map
  $\scr{B}(\scr{H}) \ni A \mapsto U (\Bbb{1} \otimes A) U^* \in \scr{B}(\scr{H})$ is an operation with the required properties.  
\end{proof}

As an immediate corollary we can show that ``not type I'' together with Haag duality implies infinite one copy
entanglement.

\begin{kor} \label{kor:4}
  Consider a simple bipartite system $\goth{A}, \goth{B} \subset \scr{B}(\scr{H})$ which is not split, but satisfies
  Haag duality. Each normal state $\omega$ of $\scr{B}(\scr{H})$ has infinite one copy entanglement with respect to
  $\goth{A}, \goth{B}$.
\end{kor}

\begin{proof}
  Since the split property does not hold, the two algebras $\goth{A}, \goth{B}$ are not of type I (Proposition
  \ref{prop:1}). Hence we can apply Proposition \ref{prop:9} to get a local, normal operation $\gamma:
  \scr{B}(\scr{H}_1) \to \scr{B}(\scr{H})$.  Since $\omega$ is normal, the state $\omega \circ \gamma$ of $\scr{B}(\scr{H}_1)$ is
  normal as well, and according to Proposition \ref{prop:6} it has infinite one copy entanglement. Hence, by
  definition we can find for all $\epsilon > 0$ and all $d \in \Bbb{N}$ a local operation $T : \scr{B}(\Bbb{C}^d \otimes
  \Bbb{C}^d) \to \scr{B}(\scr{H}_1)$ such   that  
  \begin{equation}
    \omega\bigl( \gamma \circ T[\kb{\chi_d}]\bigr) \geq 1 - \epsilon.
  \end{equation}
  Since $\gamma$ is local by assumption, this implies that $\omega$ has infinite one copy entanglement, as stated. 
\end{proof}

A second consequence of Proposition \ref{prop:9} concerns Bell inequalities. To state it we need the following
result from \cite{MR887998}.

\begin{prop} \label{prop:10}
  Consider a (not necessarily simple) bipartite system, consisting of the von Neumann algebras $\goth{A},
  \goth{B} \subset \scr{B}(\scr{H})$. The following two statements are equivalent:
  \begin{enumerate}
  \item 
    For every normal state $\omega$ we have $\beta(\omega) = \sqrt{2}$.
  \item 
    There is a unitary isomorphism under which
    \begin{equation} \label{eq:9}
      \scr{H} \cong \scr{H}_1 \otimes \tilde{\scr{H}},\quad \goth{A} \cong \scr{R}_{L,1} \otimes \tilde{\goth{A}},\quad \goth{B} \cong
      \scr{R}_{R,L} \otimes \tilde{\goth{B}}
    \end{equation}
    holds with appropriate von Neumann algebras $\tilde{\goth{A}}, \tilde{\goth{B}} \subset \scr{B}(\tilde{\scr{H}})$. 
  \end{enumerate}
\end{prop}

From this we get with Proposition \ref{prop:9}:

\begin{kor} \label{kor:5}
  Consider again the assumptions from Corollary \ref{kor:4}. Then each normal state $\omega$ of $\scr{B}(\scr{H})$
  satisfies $\beta(\omega) = \sqrt{2}$.
\end{kor}

\begin{proof}
  According to Proposition \ref{prop:9} we have a local, normal operation $\gamma:   \scr{B}(\scr{H}_1) \to
  \scr{B}(\scr{H})$, and $\sigma = \omega \circ \gamma$ becomes a normal state of $\scr{B}(\scr{H}_1)$. Proposition \ref{prop:10}
  implies that $\beta(\sigma) = \sqrt{2}$ holds. Hence for each $\epsilon>0$ there are operators $A_i \in \scr{R}_{L,1}$, $B_j \in
  \scr{R}_{R,1}$, $i=1,2$ satisfying $-\idty \leq A_i \leq \idty$, $-\idty \leq B_j \leq \idty$ and
  \begin{equation}
    \omega \circ \gamma (A_1(B_1+B_2)+A_2(B_1-B_2)) > \sqrt{2} - \epsilon.
  \end{equation}
  Since $\gamma$ is local and $\epsilon>0$ is arbitrary this, equation immediately implies that $\beta(\omega)=\sqrt{2}$ holds as
  stated. 
\end{proof}

Now we can summarize all our results to get the main theorem of this section:

\begin{thm} \label{thm:3}
  Consider a simple bipartite system $\goth{A}, \goth{B} \subset \scr{B}(\scr{H})$ satisfying Haag duality
  ($\goth{B} = \goth{A}'$). Then the following statements are equivalent:
  \begin{enumerate}
  \item \label{item:4}
    Each normal state on $\goth{B}(\scr{H})$ has infinite one copy entanglement.
  \item \label{item:5}
    Each separable state is singular.
  \item \label{item:8}
    The algebras $\goth{A}, \goth{B}$ are not type I.
  \item \label{item:8a}
    The split property does not hold.
  \item \label{item:12}
    Each normal state on $\scr{B}(\scr{H})$ leads to a maximal violation of Bell inequalities. 
  \item \label{item:11}
    There is a von Neumann algebra $\scr{M} \subset \scr{B}(\scr{K})$ and a unitary $U: \scr{H} \to \scr{H}_1 \otimes
    \scr{K}$ with $U \goth{A} U^* = \scr{R}_{L,1} \otimes \scr{M}$ and $ U \goth{B} U^* = \scr{R}_{R,1} \otimes \scr{M}'$.
  \item \label{item:3}
    There is a normal state on $\scr{B}(\scr{H})$ with infinite one-copy entanglement.
  \end{enumerate}
\end{thm}

\begin{proof}
  The implications \ref{item:4} $\Rightarrow$ \ref{item:5} and \ref{item:5} $\Rightarrow$ \ref{item:8} are trivial, while
  \ref{item:8} $\Rightarrow$ \ref{item:4} and \ref{item:8} $\Leftrightarrow$ \ref{item:8a} are shown in Corollary \ref{kor:4} and
  Proposition \ref{prop:1}. Hence we get \ref{item:4} $\Leftrightarrow$ \ref{item:5} $\Leftrightarrow$ \ref{item:8} $\Leftrightarrow$ \ref{item:8a}.
  
  To handle the remaining conditions note first that \ref{item:8} $\Rightarrow$ \ref{item:12} and \ref{item:3} $\Rightarrow$
  \ref{item:8} follow from Corollary \ref{kor:5} and Theorem \ref{thm:4} respectively, while \ref{item:12} $\Rightarrow$
  \ref{item:11} is a consequence of Proposition \ref{prop:10} and the fact that Haag duality holds by
  assumption. Hence it remains to show that \ref{item:3} follows from \ref{item:11}. To this end assume that
  condition \ref{item:11} holds and consider a normal state $\omega = \sigma_1 \otimes \sigma_2$ of $\scr{B}(\scr{H}_1) \otimes
  \scr{B}(\scr{K})$. According to Proposition \ref{prop:6} $\sigma_1$ (and therefore $\omega$ as well) has infinite one
  copy entanglement. Since the operation $\scr{B}(\scr{H}) \ni A \mapsto UAU^* = \gamma(A) \in \scr{B}(\scr{H}_1) \otimes
  \scr{B}(\scr{K})$ is local and normal the pull back $\omega \circ \gamma$ of $\omega$ with $\gamma$ is normal and has infinite one
  copy entanglement, which implies condition \ref{item:3}. Therefore we get the chain of equivalences
  \ref{item:8} $\Leftrightarrow$ \ref{item:12} $\Leftrightarrow$ \ref{item:11} $\Leftrightarrow$ \ref{item:3}, which concludes the proof.
\end{proof}

Hence, under the assumption of Haag duality, entanglement theory divides into two different cases: on the
one hand low entangled systems which can be described as usual in terms of tensor-product Hilbert spaces and
on the other infinitely entangled ones, which always arise if the observable algebras $\goth{A}, \goth{B}$ of
Alice and Bob are not of type I. This implies in particular that there are a lot of systems which can be
distinguished in terms of the type of the algebra $\goth{A}$ and $\goth{B}$, but not in terms of ordinary
entanglement measures (because all normal states of these systems are infinitely entangled). Nevertheless, it
seems to be likely that there are relations between  the type of $\goth{A}, \goth{B}$ and entanglement, which
go beyond the result of Theorem \ref{thm:3}. In this context it is of particular interest to look for
entanglement properties which can be associated to a whole bipartite system instead of individual
states. We come back to this discussion at the end of Section \ref{sec:local-prop}. For now, let us conclude
this Section with the remark that item \ref{item:11} of Theorem \ref{thm:3} admits an interpretation in
terms of distillation respectively dilution processes, which nicely fits into the point of view just outlined:
If we take the maximally entangled system $\scr{R}_{L/R,1}$ and add a second non-maximally entangled one ($\scr{M},
\scr{M}'$) the result $(\goth{A}, \goth{B}$) is again non-maximally entangled. Hence we have ``diluted'' the
entanglement originally contained in $\scr{R}_{L/R,1}$. If we start on the other hand with a non-maximally
entangled system $\goth{A}, \goth{B}$ and discard a lower one ($\scr{M}, \scr{M}'$) we can concentrate (or
distill) the entanglement originally contained in $\goth{A}, \goth{B}$ and get a maximally entangled system
$\scr{R}_{L/R,1}$.  

\section{Entangled spin chains}
\label{sec:entangl-spin-chains}

Let us return now to spin chains and to the C*-algebras $\scr{A}_L, \scr{A}_R \subset \scr{A}$ defined in Section
\ref{sec:preliminaries}. If $\omega$ is a pure state on the quasi-local algebra $\scr{A}$, the pair of von 
Neumann algebras $\scr{R}_{L,\omega}, \scr{R}_{R,\omega}$ form a simple bipartite system (cf. Section
\ref{sec:preliminaries}). According to Lemma \ref{lem:1} $\omega$ has infinite one copy entanglement with respect
to $\scr{A}_L, \scr{A}_R$ iff the GNS vacuum has the same property with respect to $\scr{R}_{L,\omega},
\scr{R}_{R,\omega}$. Hence we get the following simple corollary of Theorem \ref{thm:3}.

\begin{kor} \label{kor:2}
  Consider a pure state $\omega \in \scr{A}^*$ which satisfies Haag duality, i.e. $\scr{R}_{R,\omega} =
  \scr{R}_{L,\omega}'$. It has infinite one copy entanglement iff the von Neumann algebras $\scr{R}_{L/R,\omega}$ are
  not of type I.  
\end{kor}

Applying again Theorem \ref{thm:3} and Lemma \ref{lem:1} we see in addition that (under the same assumption as
in Corollary \ref{kor:2}) each $\pi_\omega$-normal state $\sigma$ has infinite one copy entanglement as well. This fact 
has a simple but interesting consequence for the stability of infinite entanglement under time evolution. To
explain the argument consider a completely positive map $T: \scr{A} \to \scr{A}$ which is  $\pi_\omega$-normal,
i.e. there is a normal cp-map $T_\omega: \scr{B}(\scr{H}_\omega) \to \scr{B}(\scr{H}_\omega)$ such that $\pi_\omega\bigl(T(A)\bigr) =
T_\omega\bigl(\pi_\omega(A)\bigr)$. Obviously, this $T$ maps $\pi_\omega$- normal states to $\pi_\omega$-normal states. Hence we get  

\begin{kor}
  Consider again a pure state $\omega \in \scr{A}^*$ which satisfies Haag duality, and a $\pi_\omega$-normal cp map $T:
  \scr{A} \to \scr{A}$. The image $T^*(\omega)$ of $\omega$ under $T$ has infinite one copy entanglement iff $\omega$ has. 
\end{kor}

We can interpret this corollary in terms of decoherence: Infinite one copy entanglement of a state $\omega$ is
stable under each decoherence process which can be described by a $\pi_\omega$-normal, completely positive time
evolution. By the same reasoning, it is impossible to reach a state with infinite one copy entanglement by a
normal operation, if we start from a (normal) separable state. This might look surprising at a first glance,
however, the result should not be overestimated: It does not mean that infinite one copy entanglement can not
be destroyed, instead the message is that operations which are normal with respect to the GNS-representation
of the initial state are too tame to describe physically realistic decoherence processes. 

\subsection{Translational invariance}
\label{sec:transl-invar}

After these general remarks, let us  have now a closer look on those properties which uses explicitly the net
structure $\Bbb{Z} \supset \Lambda \mapsto \scr{A}_\Lambda \subset \scr{A}$, which defines the kinematics of a spin chain. One of the most
important properties derived from this structure is \emph{translational invariance}. If a state $\omega$ is 
translationally invariant, we can restrict the possible types for the algebras $\scr{R}_{R/L,\omega}$ significantly,
as the following proposition shows.

\begin{prop} \label{prop:11}
  If $\omega$ is a translationally invariant pure state, the half-chain algebra $\scr{R}_{L,\omega}$ (respectively
  $\scr{R}_{R,\omega}$) is infinite, i.e. not of type II$_1$ or I$_n$ with $n < \infty$.
\end{prop}

\begin{proof}
  We only consider $\scr{R}_{L,\omega}$ because $\scr{R}_{R,\omega}$ can be treated similarly. Assume that
  $\scr{R}_{L,\omega}$ is a finite factor. Then there is a (unique) faithful, normal, tracial state $\tilde{\psi}$ on
  $\scr{R}_{L,\omega}$, which gives rise to a state $\psi = \tilde{\psi} \circ \pi_\omega$ on $\scr{A}_L$. Obviously $\psi$ is
  factorial and quasi-equivalent to the restriction of $\omega$ to $\scr{A}_L$. Hence by Corollary  2.6.11 
  of \cite{BraRob1} we find for each $\epsilon > 0$ an $n \in -\Bbb{N}$ such that $|\omega(Q) - \psi(Q)| < \epsilon/2 \|Q\|$ holds for
  all $Q \in \scr{A}$ which are located in the region $(-\infty,n]$. Now consider $A,B \in \scr{A}_{[0,k]}$ for some $k
  \in \Bbb{N}$ with $\|A\| = \|B\| = 1$. Then we get with $j > n+k$ and due to translational invariance  
  \begin{equation}
    |\omega(AB) - \psi\bigl(\tau_{-j}(AB)\bigr)| = |\omega\bigl(\tau_{-j}(AB)\bigr) -  \psi\bigl(\tau_{-j}(AB)\bigr)| < \epsilon/2.
  \end{equation}
  Hence
  \begin{equation}
    |\omega(AB) - \omega(BA)| \leq |\omega(AB) - \psi\bigl(\tau_{-j}(AB)\bigr)| + | \psi\bigl(\tau_{-j}(AB)\bigr) - \omega(BA)| < \epsilon. 
  \end{equation}
  Since $\epsilon$ and $k$ were arbitrary we get $\omega(AB) = \omega(BA)$ for all $A, B \in \scr{A}_{loc}$ and by continuity
  for all $A, B \in \scr{A}$. Hence $\omega$ is a tracial state on $\scr{A}$ which contradicts the assumption that $\omega$
  is pure.   
\end{proof}

We do not yet know whether even more types can be excluded. However, the only cases where concrete examples
exist are I$_\infty$ (completely separable states of the form $\phi^{\otimes \Bbb{Z}}$) and III$_1$ (the critical XY-model
with $\gamma=0$; cf. Section \ref{sec:ground-state}). Our conjecture is that these are the only possibilities. 

Another potential simplification arising from translational invariance concerns Haag duality. We expect that
each translationally invariant pure state automatically satisfies Haag duality. However, we are not yet able to
prove this conjecture. If it is true we could replace Haag duality in Corollary \ref{kor:2} by translational
invariance, which is usually easier to test (in particular if $\omega$ is the ground state of a translationally
invariant Hamiltonian). 

Finally, note that we can discuss all these question on a more abstract level, because we only need the
unitary $V: \scr{H}_\omega \to \scr{H}_\omega$ which implements the shift $\tau$, in addition to the bipartite system
$\scr{R}_{L/R,\omega}$. All other (local) algebras can be reconstructed by
\begin{equation} 
  \scr{A}_0 = V R_{L,\omega} V^* \cap R_{R,\omega},\quad \scr{A}_j = V^j \scr{A}_0 V^{-j},
\end{equation}
and appropriate products of the $\scr{A}_j$. 

\subsection{Localization properties}
\label{sec:local-prop}

The message of Theorem \ref{thm:3} and Corollary \ref{kor:2} is that whenever we have a spin chain in a pure
state $\omega$, satisfying Haag duality (or a state quasi-equivalent to such an $\omega$) we can generate as much
singlets as we want by operations which are located somewhere in the left and right half-chains
respectively. However, these localization properties can be described a little bit more precise. To this end
let us introduce the following definition: 

\begin{defi}
  Consider two regions $\Lambda_1, \Lambda_2 \subset \Bbb{Z}$ with $\Lambda_1 \cap \Lambda_2 = \emptyset$. An operation $T: \scr{B}(\Bbb{C}^d \otimes
  \Bbb{C}^d) \to \scr{A}$ is \emph{localized} in $\Lambda_1$ and $\Lambda_2$ if $T$ is local in the sense of Definition
  \ref{def:2} and if $T\bigl(\scr{B}(\Bbb{C}^d) \otimes \Bbb{1}\bigr) \subset \scr{A}_{\Lambda_1}$ and $T\bigl( \Bbb{1} \otimes
  \Bbb{B}(\Bbb{C}^d)\bigr) \subset \scr{A}_{\Lambda_2}$ holds. 
\end{defi}

\begin{thm} \label{thm:2}
  Consider a pure state $\omega$ on $\scr{A}$, which satisfies Haag duality and which has infinite one-copy
  entanglement. Then the following statement hold: For all $\epsilon>0$, $M \in -\Bbb{N}$, $N \in [-M, \infty)$ and $d \in
  \Bbb{N}$ we can find an operation $T$ which is localized in $(-\infty,M)$ and $[M+N,\infty)$ and which satisfies
  $\omega\bigl(T(\kb{\chi_d})\bigr)> 1 - \epsilon$.  
\end{thm}

\begin{proof}
  Without loss of generality we can assume $M = 0$, because the proof is easily adopted to general $M$ (by
  translating $\omega$ appropriately). In addition let us denote the region $[0,N)$ by $\Lambda$ and set $\Lambda^c = \Bbb{Z} \setminus
  \Lambda$. Since $\scr{R}_{\Lambda,\omega} = \pi_\omega(\scr{A}_\Lambda)''$ is finite dimensional, it must be of type $I$. Hence there
  are Hilbert spaces $\scr{H}_{\Lambda,\omega}$ and $\scr{H}_{\Lambda^c,\omega}$ with 
  \begin{equation}
    \scr{H}_\omega = \scr{H}_{\Lambda,\omega} \otimes \scr{H}_{\Lambda^c,\omega},\quad \scr{R}_{\Lambda,\omega} = \scr{B}(\scr{H}_{\Lambda,\omega}) \otimes \Bbb{1},\quad
    \scr{R}_{\Lambda^c,\omega} = \Bbb{1} \otimes \scr{B}(\scr{H}_{\Lambda^c,\omega}).
  \end{equation}
  Since $\scr{R}_{L,\omega}$ and $\scr{R}_{[N,\infty),\omega}$ are subalgebras of $\scr{R}_{\Lambda^c,\omega}$ they can be
  written as
  \begin{equation} \label{eq:11}
    \scr{R}_{L,\omega} = \Bbb{1} \otimes \tilde{\scr{R}}_{L,\omega},\quad \scr{R}_{[N,\infty),\omega} = \Bbb{1} \otimes \tilde{\scr{R}}_{R,\omega}
  \end{equation}
  with two von Neumann algebras $\tilde{\scr{R}}_{L/R,\omega}$ which act on $\scr{H}_{\Lambda^c,\omega}$ and which are
  isomorphic to $\scr{R}_{L,\omega}$ and $\scr{R}_{[N,\infty),\omega}$ respectively. We see immediately that $\tilde{\scr{R}}_{L,\omega}
  \lor \tilde{\scr{R}}_{R,\omega} = \scr{B}(\scr{H}_{\Lambda^c,\omega})$ follows from the corresponding property of
  $\scr{R}_{L/R,\omega}$. In addition $\tilde{\scr{R}}_{L,\omega}$ and $\tilde{\scr{R}}_{R,\omega}$ are mutually commuting,
  hyperfinite and $\sigma$-finite. Hence they form a simple bipartite system, as defined at the beginning of
  Section \ref{sec:entangl-von-neum}. To finish the proof we only have to show that $\tilde{\scr{R}}_{L/R,\omega}$
  are not of type I and satisfy Haag duality. The statement then follows from Theorem \ref{thm:3}. 

  Since $\omega$ has infinite one-copy entanglement $\scr{R}_{L/R,\omega}$ are according to Theorem \ref{thm:3} not of
  type I. Hence Equation (\ref{eq:11}) implies immediately that $\tilde{\scr{R}}_{L,\omega}$ can not be of type I
  either. A similar statement about $\tilde{\scr{R}}_{R,\omega}$ follows from $\scr{R}_{R,\omega} =
  \scr{B}(\scr{H}_{\Lambda,\omega}) \otimes \tilde{\scr{R}}_{R,\omega}$. To show Haag duality consider $A \in
  \tilde{\scr{R}}_{L,\omega}'$. Then we have $\Bbb{1} \otimes A \in \scr{R}_{L,\omega}' = \scr{R}_{R,\omega}$. Since $\scr{R}_{R,\omega} =
  \scr{B}(\scr{H}_{\Lambda,\omega}) \otimes \tilde{\scr{R}}_{R,\omega}$ this implies $A \in \tilde{\scr{R}}_{R,\omega}$ as
  required. Together with the previous remark this concludes the proof.
\end{proof}

It is interesting to compare this result with the behavior of other models: If we consider a quantum field and
two tangent, wedge-shaped subsets of spacetime as localization regions the vacuum state has infinite one copy
entanglement under quite general conditions \cite{MR887998}. If the regions do not touch, however, the
entanglement is finite and decays quite fast as a function of the (space-like) distance of the wedges (but
entanglement never vanishes completely \cite{MR2153773}). In a harmonic oscillator chain the entanglement is
always finite even if we consider two adjacent half-chains, and it (almost) vanishes if we tear the
half-chains apart \cite{AEPW02}. In both examples the entanglement is mainly located at the place
where the localization regions meet and is basically negligible at large distances. For a spin chain in a
state with infinite one copy entanglement it is exactly the other way round. 

At a first glance the result from Theorem \ref{thm:2} seems to be quite obvious: A finite number of qubits can
carry only a finite amount of entanglement. Subtracting a finite number from infinity remains infinite. This
argument is, however, incomplete, because it assumes implicitly that entanglement is localized along the chain,
such that ignoring a finite part in the middle can not disturb the entanglement of the rest. The following
corollary shows that this type of localization is indeed possible. 

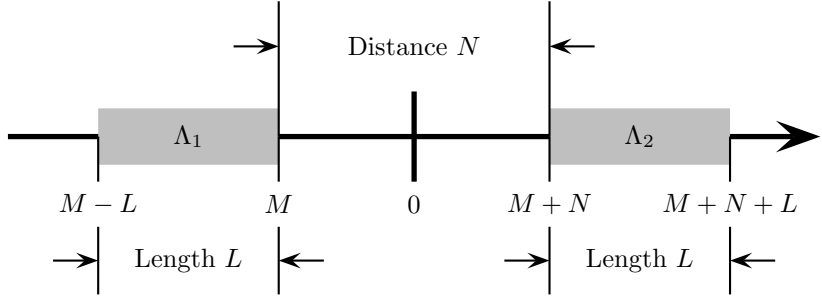
\begin{figure}[htbp]
  \begin{picture}(16,8)
    \psline[linewidth=2pt]{->}(-1,4)(17,4)
    \psline[linewidth=2pt]{-}(8,5)(8,3)
    \psline[linecolor=lightgray, linewidth=.75cm](1,4)(5,4)
    \psline[linecolor=lightgray, linewidth=.75cm](11,4)(15,4)
    \rput(3,4){$\Lambda_1$}
    \rput(13,4){$\Lambda_2$}
    \psline{-}(1,4)(1,3)
    \psline{-}(5,4)(5,3)
    \psline{-}(11,4)(11,3)
    \psline{-}(15,4)(15,3)
    \rput(8,2.5){$0$}
    \rput(1,2.5){$M-L$}
    \rput(5,2.5){$M$}
    \rput(11,2.5){$M+N$}
    \rput(15,2.5){$M+N+L$}
    \psline{-}(1,2)(1,0.5)
    \psline{-}(5,2)(5,0.5)
    \psline{-}(11,2)(11,0.5)
    \psline{-}(15,2)(15,0.5)
    \psline{->}(0,1.25)(1,1.25)
    \psline{<-}(5,1.25)(6,1.25)
    \rput(3,1.25){Length $L$}
    \psline{->}(10,1.25)(11,1.25)
    \psline{<-}(15,1.25)(16,1.25)
    \rput(13,1.25){Length $L$}
    \psline{-}(5,4)(5,7)
    \psline{-}(11,4)(11,7)
    \psline{->}(4,6)(5,6)
    \psline{<-}(11,6)(12,6)
    \rput(8,6){Distance $N$}
  \end{picture}
  \centering
  \caption{Localization regions $\Lambda_1$, $\Lambda_2$ from Corollary \ref{kor:3}.}
  \label{fig:1}
\end{figure}

\begin{kor} \label{kor:3}
  Consider the same assumptions as in Theorem \ref{thm:2}. For all $\epsilon > 0$, $M \in - \Bbb{N}$, $N \in [M,\infty)$ and
  $d \in \Bbb{N}$ there is an $L \in \Bbb{N}$ (depending in general on $N,\epsilon$ and $d$) and an operation $T$
  localized in $\Lambda_1 = [M-L,M)$ and $\Lambda_2 = [M+N,M+N+L)$ (cf. Figure \ref{fig:1}) such that
  $\omega\bigl(T(\kb{\chi_d})\bigr)> 1 - \epsilon$ holds. 
\end{kor}

\begin{proof}
  As above we can assume without loss of generality that $M=0$ holds. From Theorem \ref{thm:2}  we know that
  an operation $S: \scr{B}(\Bbb{C}^d \otimes \Bbb{C}^d) \to \scr{A}$ exists, which is localized in $(-\infty,0)$ and
  $[N,\infty)$ and which satisfies 
  \begin{equation} \label{eq:18}
    \omega(A) > 1 - \epsilon/2\ \text{with}\ A = S(\kb{\chi_d})
  \end{equation}
  The operator $A$ can be written as a limit over a net $A_\Lambda \in \scr{A}_\Lambda$, ($\Lambda \subset \Bbb{Z}$, finite), i.e. for
  each $\epsilon > 0$ there is an $\Lambda_\epsilon$ such that $\Lambda \supset \Lambda_\epsilon$ implies $\|A - A_\Lambda\| < \epsilon/4$. Now consider $\Lambda = [-L,N+L)$
  such that $\Lambda_\epsilon \subset \Lambda$ and $\Lambda^c = \Bbb{Z} \setminus \Lambda$. On $\scr{A}_{\Lambda^c}$ we can define the state $\sigma = \bigotimes_{j \in \Lambda^c}
  \sigma^{(j)}$ with $\sigma^{(j)}(B) = \tr(B)/2$ and this leads to the operation (where $\Id_\Lambda$ denotes the identity
  map on $\scr{A}_\Lambda$, and we have denoted the map $\scr{A}_{\Lambda^c} \ni A \mapsto \sigma(A) \Bbb{1} \in \scr{A}_{\Lambda^c}$ again
  with $\sigma$)   
   \begin{equation}
    \scr{B}(\Bbb{C}^d \otimes \Bbb{C}^d) \ni B \mapsto \sigma \otimes \Id_\Lambda \bigl(T(B)\bigr) \in \scr{A}_\Lambda,
  \end{equation}
  which is localized in $[-L,0]$ and $[N,N+L)$. Now note that the map $\sigma \otimes \Id_\Lambda$ is idempotent with $\|\sigma \otimes
  \Id_\Lambda\|=1$ (since $\sigma$ is a state and therefore completely positive and unital). Hence we get 
  \begin{multline}
    \|A - \sigma \otimes \Id_\Lambda (A)\| \leq \|A - A_\Lambda\| + \|A_\Lambda - \sigma \otimes \Id_\Lambda(A)\| \\ \leq \frac{\epsilon}{4} + \|\sigma \otimes \Id_\Lambda\| \|A_\Lambda - A\| \leq \frac{\epsilon}{2},
  \end{multline}
  therefore $|\omega(A - \sigma \otimes \Id_\Lambda (A))| \leq \frac{\epsilon}{2}$ and this implies with (\ref{eq:18})
  \begin{equation}
    \omega\bigl(\sigma \otimes \Id_\Lambda\bigl[S(\kb{\chi_d})\bigr]\bigr) = \omega\bigl(\sigma \otimes \Id_\Lambda(A)\bigr) \geq \omega(A) - \frac{\epsilon}{2} \geq 1 - \epsilon.  
  \end{equation}
  Hence the statement follows with $T = (\sigma \otimes \Id_\Lambda) S$. 
\end{proof}

This corollary strongly suggests the introduction of a function $L_\omega(M,N,\epsilon)$ which associates to a position $M$
and a distance $N$ the minimal length $L_\omega$ of the localization regions which is needed to extract a maximally
entangled qubit pair with accuracy $0 < \epsilon < 1$ from a chain in the state $\omega$. For a state with infinite one
copy entanglement, $L$ is well defined and always finite. Hence it provides a method to distinguish between
different states with  infinite one copy entanglement.

To get an idea what $L_\omega$ can possibly tell us about $\omega$, consider first its dependence on $\epsilon$. We can get rid
of it by defining $L_\omega(M,N) = \sup_\epsilon L_\omega(M,N,\epsilon)$. However, this quantity can become infinite if the
entanglement contained in $\omega$ is not perfectly localized (i.e. we can never extract a perfect singlet at
position $M$ and distance $N$). In this case the dependence of $L_\omega$ on $\epsilon$ is a measure of the \emph{degree
  of localization} of the entanglement contained in $\omega$. To discuss the parameters $M$ and $N$ note that 
two quasi-equivalent factor states $\omega, \sigma$ become indistinguishable ``far outside'', i.e. for each $\delta > 0$
there is a $K \in \Bbb{N}$ such that
\begin{equation}
  A \in \scr{A}_{\{|j| > K \}} \Rightarrow |\omega(A) - \sigma(A)| < \delta \|A\| 
\end{equation}
holds \cite[Cor. 2.6.11]{BraRob1}. This indicates that the asymptotic behavior of $L_\omega$ for $M \to \pm \infty$,
respectively $N \to \infty$ characterizes the \emph{folium} of $\omega$ (i.e. the equivalence class under
quasi-equivalence) while the behavior for finite $M, N$ distinguishes different states in the same
folium. (This observation matches the discussion from the end of Section \ref{sec:haag-duality}.)  In both
cases the dependence of $L_\omega$ on $M$ and $N$ describes how entanglement is distributed along the chain ($M$)
and how it decays if the distance $N$ of the localization regions grows.

Closely related to $L_\omega$ is the one-copy entanglement $E_1(\omega_\Lambda)$ of the restriction $\omega_\Lambda$ of $\omega$ to $\scr{A}_\Lambda
= \scr{A}_{\Lambda_1} \otimes \scr{A}_{\Lambda_2}$, $\Lambda = \Lambda_1 \cup \Lambda_2$, with respect to the splitting $\scr{A}_{\Lambda_1}, \scr{A}_{\Lambda_2}
\subset  \scr{A}_\Lambda$: For each $L \geq L_\omega(M,N)$ we get $E_1(\omega_\Lambda)\geq1$, if $\Lambda_1, \Lambda_2$ are disjoint regions of length $L$,
at position $M$ and with distance $N$ (cf. Figure \ref{fig:1}). This fact can be used to calculate $L_\omega(M,N)$
if we have a method to compute $E_1(\omega_\Lambda)$. Another closely related quantity is the one copy entanglement
$E_1(\omega)$ of $\omega$ with respect to the  splitting of the whole chain into a \emph{finite contiguous block} of
legth $L$ and the rest. Explicit calculation of this type are available in \cite{Eisert05a,OLEC05}, where it
is shown that $E_1$ diverges for critical chains logarithmically in $L$. Unfortunately the methods used there
are restricted to pure states, and can not be applied directly to the computation of the one copy entanglement
of $\omega_\Lambda$ with respect to the bipartite system $\scr{A}_{\Lambda_1}, \scr{A}_{\Lambda_2} \subset \scr{A}_\Lambda$ just mentioned (since
$\omega_\Lambda$ is in general mixed, even if $\omega$ is pure).  

\subsection{Cluster properties}
\label{sec:cluster-properties}

The function $L_\omega$ just introduced provides a special way to analyze the decay of correlations as a function of
the distance (of the localization regions). A different approach with the same goal is the study of cluster
properties. In this subsection we will give a (very) brief review together with a discussion of the relations
to the material presented in this paper.    

In its most simple form, the cluster property just says that correlations vanish at infinite distances, i.e. 
\begin{equation} \label{eq:32}
  \lim_{k \to \infty} \bigl| \omega\bigl(A \tau_k(B)\bigr) - \omega(A) \omega(B) \bigr| = 0
\end{equation}
should hold for all $A, B \in \scr{A}$ (this is known as the weak cluster property). This condition, however, is
to weak for our purposes, because it always holds if $\omega$ is a translationally invariant factor state
(cf. \cite[Thm. 2.6.10]{BraRob1}). Hence we have to control the decrease of correlations more carefully. One
possibility is to consider \emph{exponential clustering}, i.e. exponential decay of correlations. It is in
particular conjectured that a translationally invariant state $\omega$ satisfies the split property (cf. Section
\ref{sec:split-properties}) if 
\begin{equation}
  \left| \omega\bigl(A \tau_k(B)\bigr) - \omega(A) \omega(B) \right| \leq C(A,B) e^{-M k} \quad \forall A \in \scr{A}_L,\ B \in \scr{A}_R
\end{equation}
holds, where $C(A,B)$ is an $A, B$ dependent constant, $M$ is a positive constant (independent of $A$ and $B$)
and $k$ is any positive integer. A complete proof of this conjecture is not yet available. If it is true,
however, it would imply according to \cite{NachtSims05} that any ground state with a spectral gap (for a
Hamiltonian with finite range interaction) has the split property.   

A different, approach is to assume that the limit (\ref{eq:32}) holds (roughly speaking) uniformly in $A$. It
can be shown that this \emph{uniform} cluster property is indeed equivalent to the split property. More
precisely, the following proposition holds \cite[Prop. 2.2]{MR1828987}: 

\begin{prop} \label{prop:2}
  For each translationally invariant pure state $\omega$ on $\scr{A}$ the following two statements are equivalent.
  \begin{enumerate}
  \item 
    $\omega$ satisfies the split property, i.e. $\scr{R}_{L,\omega} \subset \scr{N} \subset \scr{R}_{R,\omega}'$ holds with a type I
    factor $\scr{N}$.
  \item 
    $\omega$ satisfies 
    \begin{equation} \label{eq:12}
      \lim_{k \to \infty} \sup_A \left| \sum_j \bigl(\omega\bigl(A_j \tau_k(B_j)\bigr) - \omega(A_j) \omega(B_j)\bigr) \right| = 0 
    \end{equation}
    where the supremum is taken over all $A \in \scr{A}_{\rm loc}$ with $\|A\| \leq 1$ and 
    \begin{equation}
      A = \sum_{j=1}^n A_jB_j,\quad A_j \in \scr{A}_R,\quad B_j \in \scr{A}_L
    \end{equation}
    for some $n \in \Bbb{N}$. 
  \end{enumerate}
\end{prop}

\section{Case study: The critical XY model}
\label{sec:critical-xy-model}

To illustrate the abstract discussion from the last two sections let us now discuss the critical XY model and
its unique ground state $\varphi_S$. To this end let us denote the GNS representation associated to $\varphi_S$ with
$(\pi_S, \scr{H}_S, \Omega_S)$ and the corresponding half-chain von Neumann algebras by $\scr{R}_{L,S}$ and
$\scr{R}_{R,S}$. The main result of this section is the following theorem which shows that the
$\scr{R}_{L/R,S}$ are not of type I and that Haag duality holds. The proof will be given in Subsection
\ref{sec:ground-state}. In addition we will provide a short review of several technical details of this model.

\begin{thm} \label{XYduality}
Consider the critical XY model (i.e. $\evl$ from Equation (\ref{eq:5}) with $\abs{\lambda} =1 , \gamma \neq 0$ or $\abs{\lambda}
<1 , \gamma = 0$). 
\begin{enumerate}
\item
  The unique ground state $\varphi_S$ is not split, i.e. $\scr{R}_{L,S}$, $\scr{R}_{R,S}$ are not of type I.
\item 
  $\varphi_S$ satisfies Haag duality
  \begin{equation} \label{eqn:a4}
    \scr{R}_{L,S}' = \scr{R}_{R,S}
  \end{equation}
\end{enumerate}
\end{thm}

According to Theorems \ref{thm:3} and \ref{thm:2} this result implies immediately that each $\pi_S$-normal state
(in particular $\varphi_S$ itself) has infinite one-copy entanglement.

\begin{kor} \label{kor:1}
  Each $\pi_S$-normal state $\omega$ on $\scr{A}$ has infinite one copy entanglement with respect to the bipartite
  system $\scr{A}_L, \scr{A}_R \subset \scr{A}$. 
\end{kor}

\subsection{The selfdual CAR algebra}
\label{sec:car-algebra}

To prove Theorem \ref{XYduality} we will use the method introduced in \cite{MR743380} by H.Araki. The idea is,
basically, to trace statements about spin chains back to statements about Fermionic systems (cf. Section
\ref{sec:jord-wign-transf}). To prepare this step we will give a short review of some material about CAR
algebras which will be used in this context. More detailed and complete presentations of this subject can be
found in \cite{MR0295702,Araki87,BaumWoll,BraRob2}.  

Hence, let us consider a complex Hilbert space $\scr{K}$ equipped with an antiunitary involution $\Gamma$. To this
pair we can associate a C*-algebra $\CAR(\scr{K},\Gamma)$ which is generated by elements $B(h) \in  \CAR (\calK
, \Gamma)$ where $h \in \scr{K}$ and $h \mapsto B(h)$ is a linear map satisfying 
\begin{equation}
  \{ B(h_1)^* ,  B(h_2) \} = (h_1,h_2)_{\scr{K}} 1, \quad B(\Gamma h)^* = B(h).
  \label{eqn:a14}
\end{equation}
$\CAR(\scr{K},\Gamma)$ is uniquely determined up to isomorphisms and called \emph{self-dual CAR algebra} over
$(\scr{K}, \Gamma)$.  If there is no risk of confusion we denote $\CAR (\scr{K} ,\Gamma)$ by $\CAR$.

Any unitary $u$ on $\calK$ satisfying $\Gamma u \Gamma =u $ gives rise to the automorphism $\beta_u$ of $\CAR$ determined by
\begin{equation} 
  \beta_u (B(h))=B(uh) .
  \label{eqn:a15}
\end{equation}
$\beta_u$ is called the Bogoliubov automorphism associated with $u$. Of particular importance is the case $u =
\Bbb{1}$ and we write
\begin{equation}
  \Theta = \beta_{-1}.
  \label{eqn:a16}
\end{equation}
$\Theta$ is an automorphism of $\CAR (\scr{K} ,J)$  specified by the following equation:
\begin{equation}
  \Theta (B(h)) = -B(h).
\end{equation}
As  the automorphism $\Theta$ is involutive, $\Theta^2 (Q)=Q$, we introduce the $\Bbb{Z}_2$ grading with respect to $\Theta$:
\begin{equation}
  \CAR_{\pm} = \{ Q \in \CAR\, | \, \Theta (Q)= \pm Q \} , \quad \CAR = \CAR_+\cup\CAR_-.
\end{equation}

Next we introduce quasi-free states of $\CAR (\scr{K}, \Gamma)$. To this end note that for each state $\psi$ of $\CAR$
there exists a bounded selfadjoint operator $A$ on the test function space $\scr{K}$ such that 
\begin{equation}
  \psi (B(h_1) B(h_2)) =  ( \Gamma h_1 , A h_2 )_{\scr{K}}
\end{equation}
and
\begin{equation}
0 \leq A \leq 1 ,  \quad \Gamma A \Gamma = 1-A .
\label{eqn:a17}
\end{equation}
holds. $A$ is called the \emph{covariance operator} for  $\psi$.

\begin{defi}
Let  $A$ be a selfadjoint operator on $\calK$ satisfying (\ref{eqn:a17}), and $\psi_A$ the state of $\CAR (\calK
,J)$ determined by 
 \begin{equation}
   \psi_A (B(h_1)B(h_2) \cdots B(h_{2n+1}) ) =0 ,
\label{eqn:a18}
\end{equation}
and
\begin{equation}
  \psi_A (B(h_1)B(h_2)\cdots B(h_{2n}) ) = \sum  sign( p ) \prod_{j=1}^{n}   ( J h_{p(2j-1)} , A h_{p(2j)} )_{\calK},
\label{eqn:a19}
\end{equation}
where the sum is taken over all permutations $p$ satisfying
\begin{equation}
  p(1) < p(3) < ... < p(2n-1) , \quad p(2j-1) < p(2j) 
\end{equation}
and $sign (p)$ is the signature of $p$. $\psi_A$ is called the \emph{quasi-free state} associated with 
the covariance operator $A$.
\end{defi}

A projection $E$ on $\scr{K}$ satisfying $\Gamma E \Gamma = 1 - E$ is called a \emph{basis projection} and the
corresponding quasi-free state $\psi_E$ is called a \emph{Fock state}. A quasi-free state is pure iff it is a Fock
state. The GNS representation $(\scr{H}_E, \pi_E, \Omega_E)$ of $\psi_E$ can be easily given in terms of the
\emph{antisymmetric Fock space} $\scr{F}_a(E\scr{K})$ over $E\scr{K}$:
\begin{equation}
  \scr{H}_E = \scr{F}_a(E\scr{K}),\quad \pi_E\bigl(B(h)\bigr) = C(EJh) + C^*(Ef),\quad \Omega_E = \Omega,
\end{equation}
where $C(f), C^*(f)$ denote annihilation and creation operators on $\scr{F}_a(E\scr{K})$ and $\Omega \in
\scr{F}_a(E\scr{K})$ is the usual Fock vacuum. 

If two quasi-free states are given we need a criterion to decide whether they are quasi-equivalent or not. This
is done by the following proposition.

\begin{prop} \label{prop:8}
  Two quasi-free states $\psi_{A_1}$, $\psi_{A_2}$ of $\CAR(\scr{K},\Gamma)$ are quasi-equivalent iff the operator
  $\sqrt{A_1} - \sqrt{A_2}$ is Hilbert-Schmidt.
\end{prop}

For two Fock states $\psi_{E_1}$, $\psi_{E_2}$ this condition reduces obviously to: $E_1 - E_2$ is Hilbert Schmidt,
and since $\psi_{E_1}$ and $\psi_{E_2}$ are pure, they are quasi-equivalent iff they are unitarily equivalent. Hence
in this case we get the statement: $\psi_{E_1}$ and $\psi_{E_2}$ are unitarily equivalent iff $E_1 - E_2$ is Hilbert
Schmidt. If only one of the two operator is a projection, Proposition \ref{prop:8} can be easily reduced to
the following statement (cf. \cite{MR0295702} for a proof):

\begin{prop} \label{prop:12}
  Consider a Fock state $\psi_E$ and a quasi-free state $\psi_A$ of $\CAR(\scr{K},\Gamma)$. They are quasi-equivalent iff
  $E - A$ and $\sqrt{A(\Bbb{1}-A)}$ are both Hilbert-Schmidt.
\end{prop}

Now consider a second projection $P$ on $\scr{K}$ and assume that $P$ commutes with $\Gamma$. Then we can define
$\CAR(P\scr{K}, P\Gamma P)$ which is a subalgebra of $\CAR(\scr{K},\Gamma)$. To state our next result (known as
``twisted duality'') concerning the commutant of the algebra  
\begin{equation}
  \scr{M}(P) = \pi_E\bigl(\CAR(P\scr{K}, P\Gamma P)\bigr)'',
\end{equation}
note that $\psi_E$ is invariant under the automorphism $\Theta$ defined in (\ref{eqn:a16}). Hence there is a unitary
$Z$ on $\scr{H}_E$ such that $\pi_E\bigl(\Theta(A)\bigr) = Z \pi_E(A) Z^*$ holds. Now we have
(cf. \cite{MR0295702,MR1915649} for a proof) 

\begin{prop}[Twisted duality] \label{prop:3}
  The von Neumann algebra 
  \begin{equation}
    \scr{N}(1-P) = \left\{ Z \pi_E\bigl(B(h)\bigr) \, | \,h \in (\Bbb{1} - P) \scr{K} \right\}''
  \end{equation}
  coincides with the commutant of $\scr{M}(P)$, i.e. $\scr{M}(P)' = \scr{N}(1-P)$ holds.
\end{prop}
 
\subsection{The Jordan Wigner transformation}
\label{sec:jord-wign-transf}

Now we will use the arguments in \cite{MR743380} to relate spin chains to Fermionic systems. The first step is to
enlarge the algebra $\calA$ to another algebra $\tilde{\calA}$ by adding a new selfadjoint unitary element $T$
which has the following property:
\begin{equation}
  T^2=1,\quad  T^* =T,\quad TQT = \Theta_-(Q) \qquad \text{for $Q$ in $\calA$,}
  \label{eqn:a5}
\end{equation}
where $\Theta_-$ is an automorphism of $\calA$ defined by
\begin{equation}
  \Theta_-(Q)  = \lim_{N \to -\infty} \rbk{ \prod_{j=-N}^{-1} \sigma_z^{(j)}} Q \rbk{ \prod_{j=-N}^{-1} \sigma_z^{(j)}} .
\label{eqn:a6}
\end{equation}
$\tilde{\calA}$ is the crossed  product by the $Z_2$ action via $\Theta_-$. Obviously
\begin{equation}
  \tilde{\calA} =\calA \cup \calA T 
\end{equation}
and we extend $\Theta_-$ to $\tilde{\scr{A}}$ by $\Theta_-(T) = T$.

We introduce another automorphism $\Theta$ via the formula,
\begin{equation}
  \Theta (Q)  = \lim_{N \to \infty} \rbk{ \prod_{j=-N}^{N} \sigma_z^{(j)}} Q \rbk{ \prod_{j=-N}^{N} \sigma_z^{(j)}} .
  \label{eqn:a7}
\end{equation}
Thus
\begin{equation}
  \Theta (\sigma_x^{(j)}) = - \sigma_x^{(j)},\quad   \Theta (\sigma_y^{(j)}) = - \sigma_y^{(j)},\quad   \Theta (T) = T,
\end{equation}
and we set
\begin{equation}
  \calA_{\pm} = \cbk{ Q \in \calA \,|\, \Theta (Q) = \pm Q } .
\end{equation}

Now  we can realize the creation and annihilation operators of fermions in $\tilde{\calA}$ as follows.
\begin{equation}
  c^*_j = TS_j ( \sigma_x^{(j)}+ i \sigma_y^{(j)} )/2, \quad  c_j = TS_j ( \sigma_x^{(j)}- i \sigma_y^{(j)} )/2
  \label{eqn:a8}
\end{equation}
where
\begin{equation}
  S_j = \begin{cases}
      \sigma_z^{(0)}\cdots\sigma_z^{(j-1)} & \text{ for $j \geq 1$, }\\ 
      1 & \text{ for $j =0$,}\\
      \sigma_z^{(-j)}\cdots\sigma_z^{(-1)} & \text{ for $j \leq -1$. }\\ 
    \end{cases}
  \label{eqn:a9}
\end{equation}
Operators $c^*_j $ and $c_j $ satisfy
the canonical anticommutation relations (\ref{eqn:a10}).
\begin{equation}
  \{ c_j , c_k \} =\{ c^*_j , c^*_k \}=0, \quad 
  \{ c_j , c^*_k \}= \delta_{j,k} 1 
  \label{eqn:a10}
\end{equation}
for any integer $j$ and $k$. 

 For a vector $f = (f_j) \in l_2(\Bbb{Z}) $, we set 
\begin{equation}
   c^*(f) = \sum_{j \in \Bbb{Z}}  c_j^* f_j, \quad c(f) = \sum_{j \in \Bbb{Z}}  c_j f_j 
   \label{eqn:a11} 
\end{equation}
where the sum converges in norm topology of $\tilde{\calA}$. Furthermore, let
\begin{equation}
  B(h)= c^*(f_1) + c(f_2) 
  \label{eqn:a12}
\end{equation}
where $h=(f_1\oplus f_2)$ is a vector  in the test function space $\calK = l_2({\Bbb{Z}})\oplus l_2({\Bbb{Z}})$ . 

By $\overline{f}$ we denote the complex conjugate $ \overline{f}=(\overline{f}_j) $  of $f \in l_2(\Bbb{Z})$ and we
introduce an antiunitary involution $\Gamma$ on the test function  space $\calK = l_2({\Bbb{Z}})\oplus l_2({\Bbb{Z}})$
determined by 
\begin{equation}
  \Gamma (f_1\oplus f_2) =(\overline{f}_2 \oplus \overline{f}_1) .
  \label{eqn:a13}
\end{equation}
It is easy to see that 
\begin{equation}
  \{ B(h_1)^* ,  B(h_2) \} = (h_1,h_2)_{\calK} 1, \quad B(\Gamma h)^* = B(h) .
\end{equation}
holds. Hence the elements $B(h)$ just defined generate a subalgebra of $\tilde{\scr{A}}$ which is isomorphic
to the CAR algebra $\CAR(\scr{K},\Gamma)$, and which is therefore identified with the latter. In this context note
that the two definitions of the automorphism $\Theta$ in Equation (\ref{eqn:a7}) and (\ref{eqn:a16}) are
compatible. The relation between the CAR algebra $\CAR$ and the spin chain algebra $\scr{A}$ is now given by
the following equation:
\begin{equation} \label{eq:35}
  \scr{A}_+  = \CAR_+ , \quad  \scr{A}_- = \CAR_- T ,
\end{equation}
i.e. the even parts of both algebras coincide. Note that this implies in particular that $\scr{A}$ is
generated by elements $B(h) T$ with $h \in \scr{K}$. Furthermore, the automorphisms $\tau$ and $\Theta_-$ can be
implemented as well in terms of Bogolubov transformations, provided the shift $\tau$ is extended to
$\tilde{\scr{A}}$ by  
\begin{equation}
  \tau_1 (c_{j} )= c_{j+1}  , \quad \tau_1 (c_{j}^* )= c_{j+1}^* , \quad  \tau_1 (T) = T\sigma_z^{(0)} = T (2 c_{0}^* c_{0} -1 )
  \label{eqn:a20}
\end{equation}
Now we define for $f = (f_j) \in l_2({\Bbb{Z}})$ the operators
\begin{equation}
  (uf)_j =  f_{j-1} ,
  \label{eqn:a21}
\end{equation}
and
\begin{equation}
  (\theta_- f)_j = 
  \begin{cases}
    f_j & \text{ for $j \geq 0$ , }\\
    - f_j & \text{ for $j \leq -1$ .}
  \end{cases}
  \label{eqn:a22}
\end{equation}
By abuse of notation, we denote operators $\theta_-$ and
$u$ on  $\calK = l_2({\Bbb{Z}})\oplus l_2({\Bbb{Z}})$ by the same symbols:
\begin{equation}
  u ( f_1 \oplus f_2 ) = ( u f_1 \oplus u f_2 ) , \quad \theta_- ( f_1 \oplus f_2 ) = ( \theta_- f_1 \oplus \theta_- f_2 ).
\end{equation}
Then we have
\begin{equation}
  \tau_1\bigl(B(h)\bigr)=B(u h), \quad  \Theta_-\bigl(B(h)\bigr) = B(\theta_- h),
  \label{eqn:a23}
\end{equation}
for all $h \in \scr{K}$.

Now we are interested in states $\omega$ on $\scr{A}$ which are $\Theta$-invariant. Since $\Theta(A) = -A$ for each $A \in
\scr{A}_-$ this implies that $\omega$ is uniquely determined by its restriction to $\scr{A}_+$. Due to Equation
(\ref{eq:35}) this restriction can arise in particular from a Fock state $\psi_E$ of $\CAR$, i.e.
\begin{equation}
  \omega(A) = \omega(A_+ + A_-) = \psi_E(A_+),\quad A_+ \in \scr{A}_+ = \CAR_+,\ A_- \in \scr{A}_-.
\end{equation}
For this special class of states we can trace Haag duality back to twisted duality (Proposition
\ref{prop:3}). To this end let us introduce the projection $p$ on $l_2({\Bbb{Z}})$  by
\begin{equation}
  p = \frac{\theta_- + \Bbb{1}}{2}
  \label{eqn:a27}
\end{equation}
or more explicitly, for f in $l_2({\Bbb{Z}})$
\begin{equation}
  (p f)_j = 
  \begin{cases}
    f_j & \text{ for $j \geq 0$, }\\
    0 & \text{ for $j \leq -1$.}
  \end{cases}
  \label{eqn:a28}
\end{equation}
On $\calK$ we then set
\begin{equation} \label{eq:40}
  P(f_1 \oplus f_2 ) =  (pf_1 \oplus pf_2 ).
\end{equation}
The operator $P$ defines the localization to the right half chain. With this notation we can state the
following result:

\begin{prop} \label{prop:4}
  Consider a $\Theta$ invariant state $\omega$ which coincides on $\scr{A}_+ = \CAR_+$ with the Fock state $\psi_E$. Then
  Haag duality holds, i.e. 
  \begin{equation} \label{eq:39}
    \scr{R}_{L,\omega} = \scr{R}_{R,\omega}'
  \end{equation}
  is satisfied.
\end{prop}

\begin{proof}
  The idea of the proof is to relate the GNS representation $(\scr{H}_\omega,\pi_\omega, \Omega_\omega)$ of $\omega$ to the GNS
  representation $(\scr{H}_E, \pi_E, \Omega_E)$ of $\psi_E$ (i.e. the Fock representation), and to apply twisted duality
  (Proposition \ref{prop:3}). Hence, let us consider the restriction of $\psi_E$ to $\scr{A}_+ = \CAR_+$. Its GNS
  representation is given by $(\scr{H}_E^+, \pi_E^+, \Omega_E)$ with
  \begin{equation}
    \pi_E^+(A) = \pi_E(A) \restr \scr{H}_E^+,\quad \scr{H}_E^+ = [\pi_E(\scr{A}_+)'' \Omega_E],\quad A \in \scr{A}_+.
  \end{equation}
  In addition, note that $\scr{A}$ can be written as the crossed product of $\scr{A}_+$ with respect to the
  $\Bbb{Z}_2$ action given by $\Ad(\sigma_x^{(0)})$. In other words each $A \in \scr{A}$ can be written in unique way
  as $A = A_0 + A_1 \sigma_x^{(0)}$ with $A_0, A_1 \in \scr{A}_+$. This implies that $\pi_\omega$ is uniquely determined by
  its action on $\scr{A}_+$ and $\sigma_x^{(0)}$. It is therefore straightforward to see that $\pi_\omega$ can be written
  as  
  \begin{gather} \label{eq:37}
    \scr{H}_\omega = \scr{H}_E^+ \otimes \scr{H}_E^+,\quad \Omega_\omega = \Omega_E \oplus 0,\quad \pi_\omega(\sigma_x^{(0)}) \xi \oplus \eta = \eta \oplus \xi,\\
    \pi_\omega(A) = \pi_E^+(A) \oplus \pi_E^+(\sigma_x^{(0)} A \sigma_x^{(0)}),\quad A \in \scr{A}_+.\label{eq:36}
  \end{gather}

  Alternatively, recall that $\scr{A}$ is generated by elements $B(h) T \in \scr{A}_-$ with $h \in \scr{K}$. Hence
  it is sufficient to calculate $\pi_\omega\bigl(B(h)T\bigr)$. To this end denote the orthocomplement of
  $\scr{H}_E^+$ by $\scr{H}_E^-$ and introduce the operators  
  \begin{equation} \label{eq:14}
    B_E^\pm(h) = \pi_E\bigl(B(h)\bigr) \restr \scr{H}_E^\mp,\quad h \in \scr{K}.
  \end{equation}
  From Equations (\ref{eqn:a18}) and (\ref{eqn:a19}) it follows immediately that the range of $B_E^\pm(h)$ is
  $\scr{H}_E^\pm$, hence 
  \begin{equation} \label{eq:16}
    \pi_E\bigl(B(h)\bigr) \xi \oplus \eta = B_E^+(h)\eta \oplus B_E^-(h)\xi,\quad  \xi \in \scr{H}_E^+,\ \eta \in \scr{H}_E^-.
  \end{equation}
  With $B(h)T = B(h)T \sigma_x^{(0)} \sigma_x^{(0)}$ we get from (\ref{eq:37}) and (\ref{eq:36})
  \begin{equation} \label{eq:38}
    \pi_\omega\bigl(B(h) T) \xi \oplus \eta = \pi_E^+\bigl(B(h) T \sigma_x^{(0)}\bigr) \eta \oplus \pi_E^+\bigl(\sigma_x^{(0)} B(h) T\bigr) \xi.
  \end{equation}
  Now note that $\sigma_x^{(0)} = T  B(h_0)$ holds with $(h_0)_j = (\delta_{j0}, \delta_{j0})$ -- this can be derived
  immediately from the definitions of $B(h)$ and $c_j, c_j^*$ in Equations (\ref{eqn:a8}) and
  (\ref{eqn:a12}). Hence we get from (\ref{eq:38})
  \begin{align}
    \pi_\omega\bigl(B(h) T) \xi \oplus \eta &= \pi_E^+\bigl(B(h) B(h_0)\bigr) \eta \oplus \pi_E^+\bigl(B(h_0) T B(h) T\bigr) \xi \\
    &= B_E^+(h) B_E^-(h_0) \eta \oplus B_E^+(h_0) B_E^-(\theta_-h) \xi 
  \end{align}
  where we have used $T^2 = \Bbb{1}$, $T B(h) T = \Theta_-\bigl(B(h)\bigr) = B(\theta_-h)$ and the fact that $T$
  commutes with $B(h_0)$; cf. the definition of $T$ and $\Theta_-$ in (\ref{eqn:a5}) and (\ref{eqn:a6}). This
  implies
  \begin{equation} \label{eq:13}
    U \pi_\omega\bigl(B(h)T\bigr) U^* \xi \oplus \kappa  = B_E^+(h) \kappa \oplus B_E^-(\theta_- h) \xi,\quad \xi \in \scr{H}_E^+,\ \kappa \in \scr{H}_E^-.
  \end{equation}
  where $U: \scr{H}_E^+ \oplus \scr{H}_E^+ \to \scr{H}_E^+ \oplus \scr{H}_E^-$ denotes the unitary given by
  \begin{equation} \label{eq:15}
    U \xi \oplus \eta = \xi \oplus B_E^-(h_0)\eta,\quad U^* \xi \oplus \kappa = \xi \oplus B_E^+(h_0) \kappa, 
  \end{equation}
  for each $\xi, \eta \in \scr{H}_E^+$ and $\kappa \in \scr{H}_E^-$.

  To continue the proof recall that $Z$ is the unitary on $\scr{H}_E$ which implements the automorphism $\Theta$ of
  $\CAR$. Hence $Z A_+ Z^* = A_+$ for $A \in \CAR_+$ and $Z A_- Z^* = -A_-$ for $A_- \in \CAR_-$. Since the even
  algebra $\CAR_+$ is generated by monomials $B(h_1) \cdots B(h_{2n})$ with an even number of factors, we see that
  $A_+ \scr{H}_E^+ \subset \scr{H}_E^+$ and $A_+ \scr{H}_E^- \subset \scr{H}_E^-$ hold for each $A_+ \in \CAR_+$. Similarly
  we have $A_- \scr{H}_E^+ \subset \scr{H}_E^-$ and vice versa if $A_- \in \CAR_-$. This implies immediately that $Z$
  is given (up to a global phase) by $Z \xi = \xi$ and $Z  \kappa= - \kappa$ for $\xi \in \scr{H}_E^+$ and $\kappa \in
  \scr{H}_E^-$. Since $\theta_-(Ph) = Ph$ and $\theta_-([\Bbb{1}-P]h) = - [\Bbb{1} - P]h$ hold, we get from
  (\ref{eq:13})
  \begin{gather}
    U \pi_\omega\bigl(B(Ph) T\bigr) U^* = \pi_E\bigl(B(Ph)\bigr),\\
    U\pi_\omega\bigl(B([\Bbb{1}-P]h) T\bigr)U^* = Z \pi_E\bigl(B([\Bbb{1} - P]h)\bigr).
  \end{gather}
  In addition we have 
  \begin{gather}
    \scr{R}_{L,\omega} = \bigl\{ \pi_\omega\bigl(B([\Bbb{1}-P]h)T\bigr) \, | \, h \in \scr{K} \bigr\}'',\\ \scr{R}_{R,\omega} =
    \bigl\{ \pi_\omega\bigl(B(Ph)T\bigr) \, | \, h \in \scr{K} \bigr\}''. 
  \end{gather}
  Hence we get (\ref{eq:39}) from Proposition \ref{prop:3}.
\end{proof}

\subsection{The ground state}
\label{sec:ground-state}
 
Now let us return to the XY model and its ground state (cf. \cite{MR810491} for details). Recall that
the shift is defined on $\CAR$ by a Bogolubov transformation with respect to the unitary $u$ given in Equation
(\ref{eqn:a21}). A quasi-free state  $\psi_A$ is translationally invariant if and only if the covariance operator
$A$ commutes with this $u$. It turns out that for a translationally invariant quasi-free state  $\psi_A$, the
Fourier transform $FAF^{-1}$ of the covariance operator $A$ is  a (2 by 2 matrix valued) multiplication
operator $\tilde{A}(x)$ on $F{\calK} = L_2([0, 2\pi]) \oplus L_2([0,2\pi]) $. We use the following normalization for
the  Fourier transform: 
\begin{equation} 
  F(f)(x)=\sum_{n =- \infty}^{\infty}  e^{inx} f_n ,\quad  f_n = {(2\pi )}^{-1} \int_0^{2\pi} e^{-inx} F(f)(x)  dx  
  \label{eqn:a24}
\end{equation}
for $f=(f_n) \in l_2(\Bbb{Z})$ and $F(f)(x) \in L_2([0,2\pi]) $. The $\Theta$ invariant ground state of the XY model
$\varphi_S$ is described by   
\begin{equation}
  \varphi_S (Q) = \varphi_S (Q_+ + Q_-) = \psi_E (Q_+),
  \label{eqn:a25}
\end{equation}
where $Q=Q_+ + Q_-$, $Q_{\pm } \in \calA_{\pm}$, and $E$ is the basis projection defined by the multiplication
operator  on $F\calK$; 
\begin{equation}
  F E F^{-1}= \hat{E}(x) = \frac{1}{2}\left( 1+ \frac{1}{k(x)}K(x) \right) \label{eqn:a26}
\end{equation}
with
\begin{equation}
  K(x) = \left[ \begin{array}{cc} \cos x - \lambda &- i \gamma \sin x \\ i \gamma \sin x & - (\cos x - \lambda )
    \end{array}\right], 
\end{equation}
and
\begin{equation}
  k(x)= [(\cos x - \lambda)^2 + \gamma^2 \sin^2x]^{1/2}.
\end{equation}
We will denote the GNS representation of $\varphi_S$ by $(\scr{H}_S, \pi_S, \Omega_S)$ and the left/right half-chain
algebras by $\scr{R}_{L/R,S}$. From Proposition \ref{prop:4} we immediately get:

\begin{kor}
  The unique ground state $\varphi_S$ of the critical XY model satisfies Haag duality, i.e.
  \begin{equation}
    \scr{R}_{L,S} = \scr{R}_{R,S}'
  \end{equation}
  holds.
\end{kor}

The next step is to analyze the type of the half-chain algebras $\scr{R}_{L/R,S}$. For an isotropic chain
($\gamma=0$) with magnetic field $|\lambda|< 1$ this is done in \cite[Thm. 4.3]{MR1828987} using methods from
\cite{MR1645078} 

\begin{prop}
  Consider the ground state $\varphi_S$ in the special case $\gamma=0, |\lambda|<1$. Then the von Neumann algebras
  $\scr{R}_{R/L,S}$ are of type III$_1$.
\end{prop}

In the general case we are not yet able to prove such a strong result. We can only show that the
$\scr{R}_{L/R,S}$ are not of type I (as stated in Theorem \ref{XYduality}). This is done in a series of steps,
which traces the problem back to a statement about quasi-inequivalence of quasi-free states.

\begin{lem} \label{lem:10}
  Consider a pure state $\omega$ on $\scr{A}$ and its restrictions $\omega_{L/R}$ to $\scr{A}_{L/R}$. Assume that the
  von Neumann algebras $\scr{R}_{L/R,\omega}$ are of type I, then $\omega$ and $\sigma = \omega_L \otimes \omega_R$ are
  quasi-equivalent and factorial.
\end{lem}

\begin{proof}
  Since $\scr{R}_{R,\omega}$ and $\scr{R}_{L,\omega}$ are of type I, we can decompose the GNS Hilbert space
  into a tensor product $\scr{H}_\omega = \scr{H}_{L,\omega} \otimes \scr{H}_{R,\omega}$ with $\scr{R}_{R,\omega} = \Bbb{1} \otimes
  \scr{B}(\scr{H}_{R,\omega})$ and $\scr{R}_{L,\omega} = \scr{B}(\scr{H}_{L,\omega}) \otimes \Bbb{1}$. The state $\sigma = \omega_L \otimes \omega_R$ is
  $\omega$-normal and it can be written as $\sigma(A) = \tr\bigl(\pi_\omega(A) \rho_L \otimes \rho_R)$ where $\rho_{L/R}$ are partial
  traces of $\kb{\Omega_\omega}$ over $\scr{H}_{R/L,\omega}$. The GNS representation of $\sigma$ is therefore given by
  $\scr{H}_\sigma = \scr{H}_S \otimes \scr{K}$ and $\pi_\sigma(A) = \pi_\omega(A) \otimes \Bbb{1}$ with an auxiliary Hilbert space
  $\scr{K}$. Hence $\pi_\sigma(\scr{A})'' = \scr{B}(\scr{H}_\omega) \otimes \Bbb{1}$ which shows that $\sigma$ is factorial. Since
  $\omega$ is factorial as well, the two states are either quasi-equivalent or disjoint, and since $\sigma$ is
  $\omega$-normal they are quasi-equivalent.  
\end{proof}

Hence, to prove that $\scr{R}_{L/R,S}$ are not of type I, we have to show that $\varphi_S$ and $\varphi_{L,S} \otimes \varphi_{R,S}$
are quasi-inequivalent. The following lemmas helps us to translate this to a statement about states on
$\CAR$. 

\begin{lem} \label{lem:4}
  Consider two $\Theta$-invariant states $\omega_1, \omega_2$ on $\scr{A}$ and their restrictions $\omega_1^+, \omega_2^+$ to the even
  algebra $\scr{A}_+$. Assume in addition that $\omega_1$ is pure and $\omega_2^+$ factorial. If  $\omega_{1}$ and
   $\omega_{2}$ are quasi-equivalent one of the following is valid:
   \begin{enumerate}
   \item 
     The restriction to the even part $\omega^{+}_{1}$ is quasi-equivalent to $\omega^{+}_{2}$.
   \item 
     The restriction to the even part $\omega^{+}_{1}$ is quasi-equivalent to $\omega^{+}_{2}\circ Ad(\sigma_{x}^{(0)})$ where
     $Ad(\sigma_{x}^{(0)})(Q) = \sigma_{x}^{(0)} Q \sigma_{x}^{(0)}$.  
   \end{enumerate}
\end{lem}

\begin{proof}
  Let us denote the GNS representation of $\omega_j^+$ by $(\scr{H}_j^+, \pi_j^+, \Omega_j^+)$ and of $\omega_j$ by
  $(\scr{H}_j, \pi_j, \Omega_j)$. Then we have with $A \in \scr{A}_+$
  \begin{gather}
    \scr{H}_j^+ = \overline{\pi_j(\scr{A}_+) \Omega_j},\quad \Omega_j^+ = \Omega_j,\notag \\
     P_j^+ \pi_j(A) P_j^+ = \pi_j^+(A)\ \text{and}\ P_j^- \pi_j(A) P_j^- = \pi_j^-(A) = \pi_j^+(\sigma_x^{(0)} A \sigma_x^{(0)}).
  \end{gather}
  where $P_j^\pm$ denote the projections onto $\scr{H}_j^+$ and its orthocomplement $\scr{H}_j^-$.
  Since $P_j^\pm \in \pi_j(\scr{A}_+)'$ the maps 
  \begin{equation}
    \pi_j(\scr{A}_+)'' \ni A \mapsto P_j^\pm A P^\pm \in \pi_j^\pm(\scr{A}_+)''
  \end{equation}
  define *-homomorphisms onto $\pi_j^\pm(\scr{A}_+)''$.  

  Now note that $\omega_1$ and $\omega_2$ are factorial. For $\omega_1$ this follows from purity (hence $\pi_1(\scr{A})'' =
  \scr{B}(\scr{H}_1)$) and for $\omega_2$ from quasi-equivalence with $\omega_1$, since the latter implies the existence
  of a *-isomorphism
  \begin{equation}
     \beta: \pi_1(\scr{A})'' \to \pi_2(\scr{A})''\ \text{with}\   \beta\bigl(\pi_1(A)\bigr) = \pi_2(A).
  \end{equation}
  Due to factoriality of $\omega_j$ the center $\scr{Z}_j$ of $\pi_j(\scr{A}_+)''$ is either trivial or
  two-dimensional. To see this, note that any operator in $\scr{Z}_j$ which commutes with $V_j =
  \pi_j(\sigma_x^{(0)})$ is in the center of $\pi_j(\scr{A})''$. Since $\omega_j$ is factorial, this implies that the
  automorphism $\pi_j(\scr{A}_+)'' \ni Q \mapsto \alpha_j(Q) = V_j Q V_j \in \pi_j(\scr{A}_+)''$ acts ergodically on $\scr{Z}_j$
  (i.e. the fixed point algebra is trivial). But $\alpha_j$ is idempotent such that each $\alpha_j(Q)Q$, $Q \in \scr{Z}_j$
  is a fixed point of $\alpha_j$. If $Q$ is a non-trivial projection this implies $\alpha_j(Q) = \Bbb{1} - Q$. By
  linearity of $\alpha_j$ this can not hold simultaneously for two orthogonal projections $Q_1, Q_2 \neq \Bbb{1} -
  Q_1$ in $\scr{Z}_j$. Hence $\scr{Z}_j$ is at most two-dimensional as stated. 

  To proceed, we have to use purity of $\omega_1$. According to Lemmas 4.1 and 8.1 of \cite{MR810491} the
  representations $\pi_1^+$ and $\pi_1^- = \pi_1 \circ \Ad(\sigma_x^{(0)})$ of $\scr{A}_+$ are irreducible and
  disjoint. Since $\pi_1^\pm(A) = P_1^\pm\pi(A)P_1^\pm$ holds for each $A \in \scr{A}_+$ the latter implies that the
  central supports $c(P_1^\pm)$ of $P_1^+$ and $P_1^- = \Bbb{1} - P_1^+$ (i.e. the smallest central projections
  in $\pi_1(\scr{A}_+)''$ containing $P_1^\pm$) are orthogonal. But this is only possible if $c(P_1^\pm) =
  P_1^\pm$. Hence $P_1^\pm$ are in the center of $\pi_1(\scr{A}_+)''$ and according to the discussion of the last
  paragraph these are the only non-trivial central projections. Applying the *-isomorphism $\beta$ we see likewise
  that $Q = \beta(P_1^+)$ and $\Bbb{1} - Q = \beta(P_1^-)$ are the only non-trivial central projections in
  $\pi_2(\scr{A}_+)''$. Since $A \mapsto P_2^+ A P_2^+$ is a *-homomorphism from $\pi_2(\scr{A}_+)''$ \emph{onto}
  $\pi_2^+(\scr{A}_+)''$ the center of $\pi_2(\scr{A}_+)''$ is mapped into the center of
  $\pi_2^+(\scr{A}_+)''$. Since $\omega_2^+$ is factorial by assumption we get $P_2^+QP_2^+ = P_2^+$ and
  $P_2^+(\Bbb{1}-Q)P_2^+ = 0$ or vice versa. This implies either $Q=P_2^+$ or $Q=P_2^-$. Hence $\beta$ maps
  $\pi_1^+(\scr{A}_+)''$ in the first case to $\pi_2^+(\scr{A}_+)''$ and in the second to
  $\pi_2^-(\scr{A}_+)''$. Therefore $\omega_1^+$ is quasi equivalent to $\omega_2^+$ or $\omega_2^+ \circ \Ad(\sigma_x^{(0)})$ as
  stated. 
\end{proof}

We will apply this lemma to states coinciding  with quasi-free states on the even part of the algebra. The
following lemmas (partly taken from \cite{MR917685,MR887223}) help us to discuss the corresponding
restrictions to $\CAR_+$. 

\begin{lem} \label{lem:6}
  Let $\omega_{1}$ and $\omega_{2}$ be quasi-free states of $\calA^{CAR}$. The restrictions to
  the even part $\omega^{+}_{1}$and $\omega^{+}_{2}$ are not quasi-equivalent, if $\omega_{1}$ and $\omega_{2}$ are not quasi-equivalent.  
\end{lem}

\begin{proof}
  cf. Proposition 1 of \cite{MR917685}.
\end{proof}

\begin{lem} \label{lem:8}
  Consider a basis-projection $E$, the covariance operator
  \begin{equation} \label{eq:23}
    F=PEP+(\Bbb{1}-P)E(\Bbb{1}-P),
  \end{equation}
  and the restrictions $\psi_E^+$, $\psi_F^+$ of the quasi-free states $\psi_E, \psi_F$ to the even algebra
  $\scr{A}_+$. If $\psi_E$ and $\psi_F$ are quasi-inequivalent, $\psi_E^+$ is quasi-inequivalent to $\psi_F^+$ and to
  $\psi_F^+ \circ \Ad(\sigma_x^{(0)})$.
\end{lem}

\begin{proof}
  Quasi-inequivalence of $\psi_E^+$ and $\psi_F^+$ follows directly from Lemma \ref{lem:6}. Hence assume $\psi_E^+$ and
  $\psi_F^+ \circ \Ad(\sigma_x^{(0)})$ are quasi-equivalent. From the proof of Proposition \ref{prop:4} recall that
  $\sigma_x^{(0)} = T B(h_0) = B(h_0) T$ holds with $h_0 \in \scr{K}$, $(h_0)_j = (\delta_{j0}, \delta_{j0})$. Therefore
  \begin{equation}
    \sigma_x^{(0)} B(h) \sigma_x^{(0)} = B(h_0) T B(h) T B(h_0) = B(h_0) B(\theta_- h) B(h_0). 
  \end{equation}
  With the anti-commutation relations (\ref{eqn:a14}) we get $\sigma_x^{(0)} B(h) \sigma_x^{(0)} = B(\vartheta h)$ with $\vartheta(h) =
  \langle h_0, \theta_-h\rangle h_0 - \theta_-h$. The operator $\vartheta$ is selfadjoint and unitary and commutes with $\Gamma$. This implies that
  $\vartheta F\vartheta$ is a valid covariance operator and $\psi_F \circ \Ad(\sigma_x^{(0)}) = \psi_{\vartheta F\vartheta}$ is therefore quasi-free. Hence by
  Lemma \ref{lem:6} quasi-equivalence of $\psi_E^+$ and $\psi_F^+ \circ \Ad(\sigma_x^{(0)})$ implies quasi-equivalence of
  $\psi_E$ and   $\psi_{F} \circ \Ad(\sigma_x^{(0)})$.  To proceed note that $\psi_F \circ \Ad(\sigma_x^{(0)})$ and $\psi_F \circ \Theta_-$ are
  unitarily equivalent. This follows immediately from $\Ad(\sigma_x^{(0)}) = \Theta_- \circ \Ad(B(h_0))$ and the fact that
  $\Ad(B(h_0))$ is an inner automorphism of $\CAR$. Therefore $\psi_E$ is quasi-equivalent to $\psi_F \circ \Theta_- =
  \psi_{\theta_-F\theta_-}$. But $\theta_- = 2 P - \Bbb{1}$ and therefore $P\theta_- = P$ and $(\Bbb{1} - P)\theta_- = (P - \Bbb{1})$
  which implies $\theta_-F\theta_- = F$. But this would imply that $\psi_E$ and $\psi_F$ are quasi-equivalent in contradiction
  to our assumption. Hence $\psi_E^+$ can not be quasi-equivalent to $\psi_F^+ \circ \Ad(\sigma_x^{(0)})$. 
\end{proof}

\begin{lem} \label{lem:3}
  Consider a quasi-free state $\psi_A$ of $\CAR$ with covariance operator $A$. Its restriction $\psi_A^+$ to the
  even algebra $\CAR_+$ is factorial if $A(\Bbb{1}-A)$ is not of trace-class.  
\end{lem} 

\begin{proof}
  cf. Proposition 2 of  \cite{MR917685}.
\end{proof}

Now consider again the ground state $\varphi_S$ and the corresponding product state $\sigma = \varphi_{S,L} \otimes \varphi_{S,R}$. On the
even algebra $\CAR_+$ they coincide with the Fock state $\psi_E$ and the quasi-free state $\psi_F$, where $E$ is the
basis projection from Equation (\ref{eqn:a26}) and $F$ is given by Equation (\ref{eq:23}). To check
quasi-equivalence we have to calculate the Hilbert-Schmidt norm of $E-F$ (cf. Proposition \ref{prop:8} and
\ref{prop:12}). Such calculations are already done in \cite{MR810491}, and we easily get the following lemma. 

\begin{lem} \label{lem:7}
  The operator 
  \begin{equation} \label{eq:25}
    X=PEP-PEPEP + (\Bbb{1}-P)E(\Bbb{1}-P) - (\Bbb{1}-P)E(\Bbb{1}-P)R(\Bbb{1}-P)
  \end{equation}
  with $E$ from Equation (\ref{eqn:a26}) is not trace-class.
\end{lem}

\begin{proof}
  According to Lemma 4.5 of \cite{MR810491} we have 
  \begin{equation}
    \|E-\theta_-E\theta_-\|_{\HS}^2 = \tr(E + \theta_-E\theta_- - E\theta_-E\theta_- -\theta_-E\theta_-E) = \infty
  \end{equation}
  Inserting $\theta_- = P - (\Bbb{1}-P)$ and using the fact that $\tr(Y) = \tr(PYP) +
  \tr\bigl((\Bbb{1}-P)Y(\Bbb{1}-Y)\bigr)$ holds for any positive operator $Y$, it is straightforward to see
  that $ \|E-\theta_-E\theta_-\|_{\HS}^2=4 \tr(X)$ holds. Hence the statement follows.
\end{proof}

Now we are ready to combine all the steps to prove that $\scr{R}_{L/R,S}$ are not of type I. The following
proposition concludes the proof of Theorem \ref{XYduality}. 

\begin{prop}
  Consider the unique ground state $\varphi_S$ of the critical XY-model and its GNS representation $(\scr{H}_S, \pi_S,
  \Omega_S)$. The half chain algebras $\scr{R}_{R,S}=\pi_S(\scr{A}_R)''$, $\scr{R}_{L,S}=\pi_S(A_L)''$ are not of type I.
\end{prop}

\begin{proof}
  Consider the operators $E, F$ and $X$ from Equations  (\ref{eqn:a26}), (\ref{eq:23}) and (\ref{eq:25}). It
  is easy to see $\|E-F\|^2_\HS = \tr(X)$. Hence $E-F$ is not Hilbert-Schmidt by Lemma \ref{lem:7} and $\psi_E$ not
  quasi-equivalent to $\psi_F$ by Proposition \ref{prop:12}. Lemma \ref{lem:8} implies therefore that $\psi_E^+$ is
  neither quasi-equivalent to $\psi_F^+$ nor to $\psi_F^+ \circ \Ad(\sigma_x^{(0)})$. The quasi-free states $\psi_E, \psi_F$
  coincides on $\CAR_+ = \scr{A}_+$ with $\varphi_S$ and $\sigma=\varphi_{S,L} \otimes \varphi_{S,R}$. In addition we know that $\varphi_S$ and
  $\sigma$ are $\Theta$-invariant, $\varphi_S$ is pure and $\sigma^+ = \psi_F^+$ is factorial. The latter follows from Lemma
  \ref{lem:3}, Lemma \ref{lem:7} and the fact that $F(\Bbb{1}-F) = X$ holds. Hence we can apply Lemma
  \ref{lem:4} to see that $\varphi_S$ and $\sigma$ are quasi-inequivalent. The statement then follows from Lemma
  \ref{lem:10}. 
\end{proof}

\section{Conclusions}
\label{sec:conclusions}

We have seen that the amount of entanglement contained in a pure state $\omega$ of an infinite quantum spin chain
is deeply related to the type of the von Neumann algebras $\scr{R}_{L/R,\omega}$. If they are of type I, the usual
setup of entanglement theory can be applied, including in particular the calculation of entanglement
measures. However, if $\scr{R}_{L/R,\omega}$ are not of type I all normal states have infinite one-copy
entanglement and all known entanglement measures become meaningless. The discussion of Section
\ref{sec:critical-xy-model} clearly shows that the critical XY model belongs to this class and it is very
likely that the same holds for other critical models. An interesting topic for future research is the
question how different states (respectively inequivalent bipartite systems) can be physically distinguished in
the infinitely entangled case. One possible approach is to look again at the von Neumann type. However,
it is very likely that additional information about the physical context is needed. A promising variant of
this idea is to look for physical condition which exclude particular cases. Proposition \ref{prop:11} is
already a result of this type and it is interesting to ask whether more types can be excluded by translational
invariance. Another possibility is to analyze localization behavior along the lines outlined at the end of
Section \ref{sec:local-prop}. In particular the asymptotics of $L_\omega$ in the limit $N \to \infty$ for a translationally
invariant state (such that $L_\omega$ does not depend on the position parameter $M$) seems to be very interesting,
because it should provide a way to characterize the folium of $\omega$ in terms of entanglement properties (cf. the
discussion in Section \ref{sec:local-prop}). A first step in this direction would be the calculation of $L_\omega$
for particular examples such as the critical XY model.  

\begin{appendix}
  
\section{Strong stability of hyperfinite type III factors}
\label{sec:strong-stab-hyperf}

The discussion in Section \ref{sec:haag-duality} relies heavily on the strong stability of hyperfinite type
III factors. While this is basically a known fact, we have not found an easily accessible
reference. Therefore, we will provide in the following a complete proof, which is based on the classification
of hyperfinite factors (cf. \cite[Ch. XIII]{MR1943007} for a detailed survey). 

Hence, let us start with a type III factor $\scr{R}$ and its continuous decomposition
\cite[Thm. XII.1.1]{MR1943006} 
\begin{equation}
  \scr{R} \cong \scr{N} \rtimes_\theta \Bbb{R},
\end{equation}
i.e. $\scr{N}$ is a type II$_\infty$ von Neumann algebra (acting on a Hilbert space $\scr{H}$), admitting a
faithful, semifinite, normal trace $\tau$, and $\theta$ is a centrally ergodic flow on $\scr{N}$ which scales $\tau$
(i.e. $\tau \circ \theta_s = e^{-s} \tau$).  The covariant system $(\scr{N}, \Bbb{R}, \theta)$ is uniquely determined (up to
conjugation) by the isomorphism class of $\scr{R}$. Therefore the central system $(\scr{Z}(\scr{N}),\Bbb{R},
\theta)$ -- the \emph{flow of weights} -- is unique as well.  

Now, consider a (hyperfinite) type II$_1$ factor $\scr{M}$ (acting on $\scr{K}$). The tensor product $\scr{R}
\otimes \scr{M}$ is type III again and satisfies
\begin{equation} \label{eq:19}
  \scr{R} \otimes \scr{M} \cong (\scr{N} \otimes \scr{M}) \rtimes_{\theta \otimes \Id} \Bbb{R}.
\end{equation}
To prove this equation, note that the crossed product on the right hand side is a von Neumann algebra acting
on the Hilbert space $\Lz(\scr{H} \otimes \scr{K}, \Bbb{R}, dx) = \Lz(\scr{H},\Bbb{R},dx) \otimes \scr{K}$ and generated
by $\pi_0(\scr{N} \otimes \scr{M})$ and $\lambda(\Bbb{R})$, where $\pi_0$ and $\lambda$ are representations of $\scr{N} \otimes \scr{M}$
and $\Bbb{R}$ respectively. They are given by 
\begin{equation} \label{eq:21}
  \bigl(\pi_0(A \otimes B) \xi\bigr)(s) = \bigl(\theta_s^{-1}(A) \otimes B\bigr) \xi(s),\quad \bigl(\lambda(t) \xi\bigr)(s) = \xi(t-s),
\end{equation}
where $A \in \scr{N}$, $B \in \scr{M}$ and $\xi \in \Lz(\scr{H} \otimes \scr{K}, \Bbb{R}, dx)$. If we set $\xi = \eta \otimes \zeta$ with
$\eta \in \Lz(\scr{H}, \Bbb{R}, dx)$ and $\zeta \in \scr{K}$ this leads to 
\begin{equation} \label{eq:22}
  \pi_0(A \otimes B) \eta \otimes \zeta = \tilde{\pi}_0(A) \eta \otimes B \zeta,\quad \lambda(t) \eta \otimes \zeta = \tilde{\lambda}(t) \eta \otimes \zeta
\end{equation}
where $\tilde{\pi}_0$ and $\tilde{\lambda}$ are the representations of $\scr{N}$ and $\Bbb{R}$ given by
\begin{equation}
  \bigl(\tilde{\pi}_0(A) \eta\bigr)(s) = \theta_s^{-1}(A) \eta(s),\quad \bigl(\tilde{\lambda}(t) \eta\bigr)(s) = \eta(t-s).
\end{equation}
But $\tilde{\pi}_0(\scr{N})$ and $\tilde{\lambda}(\Bbb{R})$ generate $\scr{N} \rtimes_\theta \Bbb{R} \cong \scr{R}$. Hence
Equation (\ref{eq:19}) follows from (\ref{eq:22}). 

Since $\scr{R}$ is a type III and $\scr{M}$ a type II factor, the tensor product $\scr{R} \otimes \scr{M}$ is again
a type III factor. If we consider in addition the (unique) tracial state $\tau_0$ on $\scr{M}$ we see that $\theta \otimes
\Id$ scales $\tau \otimes \tau_0$. Therefore Equation (\ref{eq:19}) is the continuous decomposition of $\scr{R} \otimes \scr{M}$. 

Now, let us have a look at the flow of weights associated to $\scr{R} \otimes \scr{M}$. Since $\scr{M}$ is a factor
the center of $\scr{N} \otimes \scr{M}$ coincides with $\scr{Z}(\scr{N}) \otimes \Bbb{1}$. Hence the central covariant
systems $(\scr{Z}(\scr{N}), \Bbb{R}, \theta)$ and $(\scr{Z}(\scr{N} \otimes \scr{M}), \Bbb{R}, \theta \otimes \Id)$ are mutual
conjugate. If $\scr{R}$ is hyperfinite, this fact can be used to show strong stability. To this end note first
that $\scr{R} \otimes \scr{M}$ is hyperfinite as well, because $\scr{M}$ is hyperfinite by assumption. Therefore
we can use classification theory and get three different cases:

\begin{itemize}
\item 
  $\scr{R}$ is of type III$_\lambda$ with $0 < \lambda < 1$. In this case the flow of weights of $\scr{R}$ is periodic
  with period $-\ln \lambda$. Since $(\scr{Z}(\scr{N}), \Bbb{R}, \theta)$ and $(\scr{Z}(\scr{N} \otimes \scr{M}), \Bbb{R}, \theta \otimes
  \Id)$ are conjugate the same holds for $\scr{R} \otimes \scr{M}$, i.e. $\scr{R} \otimes \scr{M}$ is type III$_\lambda$ with
  the same $\lambda$ (cf. \cite[Def. XII.1.5, Thm. XII.1.6]{MR1943006}). Strong stability ($\scr{R} \otimes \scr{M} \cong \scr{R}$)
  therefore follows from the uniqueness of hyperfinite III$_\lambda$ factors with $0 < \lambda < 1$. (cf. 
  \cite[Thm. XVIII.1.1]{MR1943007}). 
\item 
  $\scr{R}$ is of type III$_1$. Hence the center of $\scr{N}$ is trivial and since $\scr{M}$ is a factor the
  same holds for $\scr{Z}(\scr{N} \otimes \scr{M})$ -- in other words $\scr{R} \otimes \scr{M}$ is type III$_1$ again
  (cf. \cite[Def. XII.1.5, Thm. XII.1.6]{MR1943006}). Now we can proceed as above, if we use the uniqueness of
  the hyperfinite type III$_1$ factor \cite[Thm. XIII.4.16]{MR1943007}.
\item 
  $\scr{R}$ is of type III$_0$. In this case strong stability follows directly from the fact that two
  hyperfinite III$_0$ factors are isomorphic iff the corresponding flows of weights are conjugate
  \cite[Thm. XVIII.2.1]{MR1943007}. 
\end{itemize}

This list covers all possibilities and therefore the strong stability property used in the proof of
Proposition \ref{prop:9} is shown. 

\end{appendix}

\section*{Acknowledgment}
This research of M.~K. is partially supported by the Ministero Italiano dell'Universit\`a e della Ricerca (MIUR)
through FIRB (bando 2001) and PRIN 2005 and that of T.~M. by the Center of Excellence Program, Graduate School
Mathematics, Kyushu University,Japan.

\bibliographystyle{mk}
\bibliography{qinf}

\begin{thebibliography}{10}

\bibitem{MR0295702}
H.~Araki.
\newblock \emph{On quasifree states of {${\rm CAR}$} and {B}ogoliubov
  automorphisms}.
\newblock Publ. Res. Inst. Math. Sci. \textbf{6}, 385--442 (1970/71).

\bibitem{MR743380}
H.~Araki.
\newblock \emph{On the {$XY$}-model on two-sided infinite chain}.
\newblock Publ. Res. Inst. Math. Sci. \textbf{20}, no.~2, 277--296 (1984).

\bibitem{Araki87}
H.~Araki.
\newblock \emph{Bogoliubov automorphisms and fock representations of canonical
  anticommutation relations}.
\newblock In \emph{Operator algebras and mathematical physics (Iowa City, Iowa,
  1985)} ( 62, editor), Contemp. Math., pages 23--41. Amer. Math. Soc.,
  Providence, RI (1987).

\bibitem{MR810491}
H.~Araki and T.~Matsui.
\newblock \emph{Ground states of the {$XY$}-model}.
\newblock Comm. Math. Phys. \textbf{101}, no.~2, 213--245 (1985).

\bibitem{AraWoo}
H.~Araki and E.J. Woods.
\newblock \emph{A classification of factors}.
\newblock Publ. R.I.M.S, Kyoto Univ. \textbf{4}, 51--130 (1968).

\bibitem{AEPW02}
K.~Audenaert, J.~Eisert, M.B. Plenio and R.F. Werner.
\newblock \emph{Entanglement properties of the harmonic chain}.
\newblock Phys. Rev. A \textbf{66}, 042327 (2002).

\bibitem{MR1915649}
H.~Baumg{\"a}rtel, M.~Jurke and F.~Lled{\'o}.
\newblock \emph{Twisted duality of the {CAR}-algebra}.
\newblock J. Math. Phys. \textbf{43}, no.~8, 4158--4179 (2002).

\bibitem{BaumWoll}
H.~Baumg{\"a}rtel and M.~Wollenberg.
\newblock \emph{Causal nets of operator algebras}.
\newblock Akademie Verlag, Berlin (1992).

\bibitem{BotRez04}
A.~Botero and B.~Reznik.
\newblock \emph{Spatial structures and localization of vacuum entanglement in
  the linear harmonic chain}.
\newblock Phys. Rev. A \textbf{70}, 052329 (2004).

\bibitem{BraRob1}
O.~Bratteli and D.~W. Robinson.
\newblock \emph{Operator Algebras and Quantum Statistical Mechanics. I}.
\newblock Springer, New York (1979).

\bibitem{BraRob2}
O.~Bratteli and D.~W. Robinson.
\newblock \emph{Operator Algebras and Quantum Statistical Mechanics II}.
\newblock Springer, Berlin (1997).

\bibitem{MR2115123}
P.~Calabrese and J.~Cardy.
\newblock \emph{Entanglement entropy and quantum field theory}.
\newblock J. Stat. Mech. Theory Exp. , no.~6, 002, 27 pp. (electronic) (2004).

\bibitem{Cirelson}
B.S. Cirel'son.
\newblock \emph{Quantum generalizations of {Bell's} inequalities}.
\newblock Lett. Math. Phys. \textbf{4}, 93--100 (1980).

\bibitem{CliftHalv99}
R.~Clifton and H.~Halvorson.
\newblock \emph{Bipartite mixied states of infinite dimensional systems are
  generically nonseparable}.
\newblock Phys. Rev. A \textbf{61}, 012108 (2000).

\bibitem{MR735338}
S.~Doplicher and R.~Longo.
\newblock \emph{Standard and split inclusions of von {N}eumann algebras}.
\newblock Invent. Math. \textbf{75}, no.~3, 493--536 (1984).

\bibitem{Eisert05a}
J.~Eisert and M.~Cramer.
\newblock \emph{Single-copy entanglement in critical spin chains}.
\newblock Phys. Rev. A \textbf{72}, 042112 (2005).

\bibitem{MR2023564}
M.~Fannes, B.~Haegeman and M.~Mosonyi.
\newblock \emph{Entropy growth of shift-invariant states on a quantum spin
  chain}.
\newblock J. Math. Phys. \textbf{44}, no.~12, 6005--6019 (2003).

\bibitem{MR2194026}
S.~Farkas and Z.~Zimbor{\'a}s.
\newblock \emph{On the sharpness of the zero-entropy-density conjecture}.
\newblock J. Math. Phys. \textbf{46}, no.~12, 123301, 8 (2005).

\bibitem{HorCirLew}
P.~Horodecki, J.I. Cirac and M.~Lewenstein.
\newblock \emph{Bound entanglement for continuous variables is a rare
  phenomenon}.
\newblock quant-ph/0103076 (2001).

\bibitem{MR2131351}
A.~R. Its, B.-Q. Jin and V.~E. Korepin.
\newblock \emph{Entanglement in the {$XY$} spin chain}.
\newblock J. Phys. A \textbf{38}, no.~13, 2975--2990 (2005).

\bibitem{MR2083138}
B.-Q. Jin and V.~E. Korepin.
\newblock \emph{Quantum spin chain, {T}oeplitz determinants and the
  {F}isher-{H}artwig conjecture}.
\newblock J. Statist. Phys. \textbf{116}, no. 1-4, 79--95 (2004).

\bibitem{MR1468230}
R.~V. Kadison and J.~R. Ringrose.
\newblock \emph{Fundamentals of the theory of operator algebras. {V}ol. {II}},
  volume~16 of \emph{Graduate Studies in Mathematics}.
\newblock American Mathematical Society, Providence, RI (1997).
\newblock Advanced theory, Corrected reprint of the 1986 original.

\bibitem{MR2104889}
J.~P. Keating and F.~Mezzadri.
\newblock \emph{Random matrix theory and entanglement in quantum spin chains}.
\newblock Comm. Math. Phys. \textbf{252}, no. 1-3, 543--579 (2004).

\bibitem{MR2136600}
J.~P. Keating and F.~Mezzadri.
\newblock \emph{Entanglement in quantum spin chains, symmetry classes of random
  matrices, and conformal field theory}.
\newblock Phys. Rev. Lett. \textbf{94}, no.~5, 050501, 4 (2005).

\bibitem{InfEnt}
M.~Keyl, D.~Schlingemann and R.~F. Werner.
\newblock \emph{Infinitely entangled states}.
\newblock Quant. Inf. Comput. \textbf{3}, no.~4, 281--306 (2003).

\bibitem{Korepin04}
V.~E. Korepin.
\newblock \emph{Universality of entropy scaling in one dimensional gapless
  models}.
\newblock Phys. Rev. Lett. , no.~92, 096402 (2004).

\bibitem{MR2045771}
J.~I. Latorre, E.~Rico and G.~Vidal.
\newblock \emph{Ground state entanglement in quantum spin chains}.
\newblock Quantum Inf. Comput. \textbf{4}, no.~1, 48--92 (2004).

\bibitem{MR739630}
R.~Longo.
\newblock \emph{Solution of the factorial {S}tone-{W}eierstrass conjecture.
  {A}n application of the theory of standard split {$W\sp{\ast} $}-inclusions}.
\newblock Invent. Math. \textbf{76}, no.~1, 145--155 (1984).

\bibitem{MR917685}
T.~Matsui.
\newblock \emph{Factoriality and quasi-equivalence of quasifree states for
  {$Z\sb 2$} and {${\rm U}(1)$} invariant {CAR} algebras}.
\newblock Rev. Roumaine Math. Pures Appl. \textbf{32}, no.~8, 693--700 (1987).

\bibitem{MR887223}
T.~Matsui.
\newblock \emph{On quasi-equivalence of quasifree states of gauge invariant
  {CAR} algebras}.
\newblock J. Operator Theory \textbf{17}, no.~2, 281--290 (1987).

\bibitem{MR1828987}
T.~Matsui.
\newblock \emph{The split property and the symmetry breaking of the quantum
  spin chain}.
\newblock Comm. Math. Phys. \textbf{218}, no.~2, 393--416 (2001).

\bibitem{NachtSims05}
B.~Nachtergaele and R.~Sims.
\newblock \emph{{Lieb}-{Robinson} bound and the exponential clustering
  theorem}.
\newblock math-ph/0506030 (2005).

\bibitem{OLEC05}
R.~Orus, J.I. Latorre, J.~Eisert and M.~Cramer.
\newblock \emph{Half the entanglement in critical systems is distillable from a
  single specimen}.
\newblock quant-ph/0509023 (2005).

\bibitem{MR2115988}
I.~Peschel.
\newblock \emph{On the entanglement entropy for an {$XY$} spin chain}.
\newblock J. Stat. Mech. Theory Exp. , no.~12, 005, 6 pp. (electronic) (2004).

\bibitem{MR887998}
J.~S. Summers and R.~Werner.
\newblock \emph{Maximal violation of {B}ell's inequalities is generic in
  quantum field theory}.
\newblock Comm. Math. Phys. \textbf{110}, no.~2, 247--259 (1987).

\bibitem{SumWer95}
S.J. Summers and R.F.Werner.
\newblock \emph{On {Bell's} inequalities and algebraic invariants}.
\newblock Lett. Math. Phys. \textbf{33}, 321--334 (1995).

\bibitem{MR548728}
M.~Takesaki.
\newblock \emph{Theory of operator algebras. {I}}.
\newblock Springer-Verlag, New York (1979).

\bibitem{MR1943006}
M.~Takesaki.
\newblock \emph{Theory of operator algebras. {II}}, volume 125 of
  \emph{Encyclopaedia of Mathematical Sciences}.
\newblock Springer-Verlag, Berlin (2003).
\newblock Operator Algebras and Non-commutative Geometry, 6.

\bibitem{MR1943007}
M.~Takesaki.
\newblock \emph{Theory of operator algebras. {III}}, volume 127 of
  \emph{Encyclopaedia of Mathematical Sciences}.
\newblock Springer-Verlag, Berlin (2003).
\newblock Operator Algebras and Non-commutative Geometry, 8.

\bibitem{MR2153773}
R.~Verch and R.~F. Werner.
\newblock \emph{Distillability and positivity of partial transposes in general
  quantum field systems}.
\newblock Rev. Math. Phys. \textbf{17}, no.~5, 545--576 (2005).

\bibitem{MR1645078}
A.~Wassermann.
\newblock \emph{Operator algebras and conformal field theory. {III}. {F}usion
  of positive energy representations of {${\rm LSU}(N)$} using bounded
  operators}.
\newblock Invent. Math. \textbf{133}, no.~3, 467--538 (1998).

\bibitem{Werner89}
R.~F. Werner.
\newblock \emph{Quantum states with {Einstein-Podolsky-Rosen} correlations
  admitting a hidden-variable model}.
\newblock Phys. Rev. A \textbf{40}, no.~8, 4277--4281 (1989).

\bibitem{BEG}
R.~F. Werner and M.~M. Wolf.
\newblock \emph{Bound entangled gaussian states}.
\newblock Phys. Rev. Lett. \textbf{86}, no.~16, 3658--3661 (2001).

\bibitem{WOVC05}
M.~M. Wolf, G.~Ortiz, F.~Verstraete and J.~I. Cirac.
\newblock \emph{Quantum phase transitions in matrix product systems}.
\newblock cond-mat/0512180 (2005).

\end{thebibliography}

\end{document}